\documentclass[12pt,preprint]{aastex}
\usepackage{graphicx}
\usepackage{color}
\usepackage{rotating}

\newcommand {\be} {\begin{equation}}
\newcommand {\ee} {\end{equation}}

\newcommand {\MSun} {M_{\odot}}

\shorttitle{High-$z$ RQQs: hot semi-spherical flow.}

\shortauthors{Sobolewska  et al.}

\begin{document}

\title{High redshift radio quiet quasars - exploring the parameter space 
of accretion models. Part I: hot semi-spherical flow} 
\author{Ma{\l}gorzata A. Sobolewska}
\affil{Harvard-Smithsonian Center for Astrophysics, Cambridge, MA 02138\\
Nicolaus Copernicus Astronomical Center, 00-716 Warsaw, Poland}
\email{msobolewska@cfa.harvard.edu, malsob@camk.edu.pl}

\author{Aneta Siemiginowska}
\affil{Harvard-Smithsonian Center for Astrophysics, Cambridge, MA 02138}
\email{asiemiginowska@cfa.harvard.edu}

\and

\author{Piotr T. \.{Z}ycki} 
\affil{Nicolaus Copernicus Astronomical Center, 00-716 Warsaw, Poland}
\email{ptz@camk.edu.pl}

\begin{abstract}
Two families of models are currently considered to describe an
accretion flow onto black holes and production of the observed X-ray
radiation: (1) a standard cold accretion disk with a hot corona above
it and (2) an outer truncated accretion disk with a hot semispherical
inner flow. We compute spectra in the scenario with a
hot inner flow surrounded by a truncated accretion disk covered by a hot
corona and test the results on a sample of high
redshift ($z > 4$) quasars observed with {\it Chandra}.
We find that in order to reproduce
the ratio of optical to X-ray fluxes (the $\alpha_{\rm ox}$
parameter), the optical depth of the Comptonizing plasma has to be
rather low ($\tau = 0.02 - 0.15$ in the corona above the disk, and $\tau =
0.15 - 0.70$ in the hot inner flow). This, together with the observed X-ray photon
indices, implies either a high temperature in a thermal plasma ($kT_{\rm e} = 
90 - 500$ keV), or a nonthermal electron distribution in the plasma. We
put an upper limit on the disk truncation radius, $r_{\rm tr} \leq 40
R_{\rm S}$. The
modeled accretion rate is high, $\dot{m} > 0.2\, \dot{M}_{\rm Edd}$, which may suggest that
high-$z$ radio quiet quasars are analogs of X-ray binaries in their high or 
very high state.
\end{abstract}


\section{Introduction}

Modern X-ray observatories can detect the most distant objects in the
universe and provide high quality data 
for quasars at redshifts as high as $\sim 6$ (Becker et at.\ 2001; Brandt
et al.\ 2002b; Bechtold et al.\ 2003, hereafter B03), i.e. for
objects born in the very early universe. By comparing high- and
low-redshift sources, we may be able to determine whether these objects
evolve and if there is any particular pattern that they follow during
the evolution. This provides bases for studying the formation and
evolution of the universe.

We focus our studies on the radio quiet quasars (RQQs), which comprise
$\sim85\%-90\%$ of the quasar population (e.g., Stern et al.\ 2000). The fact
that RQQs are more common, and thus 
more typical, among the quasars makes it important to understand the
origin of their broad-band spectra. In these systems
a contribution from possible additional components related to the radio
emission should be negligible. 
On the other hand, RQQs are X-ray quieter than radio loud quasars
(RLQs) in terms of higher $\alpha_{\rm ox}$ parameter --- the spectral
index of a reference power law connecting the optical and X-ray rest
frame fluxes, for which we use the following definition: $\alpha_{\rm ox} = - \left (
\log{f_{\rm x}} - \log{f_{\rm o}} \right ) / \left ( \log{\nu_{\rm x}} -
\log{\nu_{\rm o}} \right )$, where $f_{\rm o}$ and $f_{\rm x}$ are rest
frame optical and X-ray fluxes at $\lambda = c / \nu_{\rm o} = 2500$
\AA\ and $E = h \nu_{\rm x} = 2$ keV (e.g., Zamorani et al.\ 1981;
Worrall et al.\ 1987; Laor et al.\ 1997) --- so the X-ray data of RQQs are of
worse quality than those of RLQs. 

Understanding observed spectral features in terms of theoretical model
parameters can provide important insights into physical processes
taking place in quasars, especially about the origins of the hard X-ray
radiation.  The ``standard'' model so far successful in explaining
optical/UV/X-ray spectra of active galactic nuclei (AGNs) (and those of X-ray binaries,
hereafter XRBs) contains as a crucial element an optically thick
accretion disk around a central black hole. Gravitational
energy is dissipated and thermalized in the accretion disk
(Shakura \& Sunyaev 1973). It is emitted in the form of thermal
radiation, giving a characteristic bump in the optical/UV band (``big
blue bump'') in AGN spectra (Shields 1978, Czerny \& Elvis 1987,
Siemiginowska et al.\ 1995, Koratkar \& Blaes 1999) or soft X-rays in
the case of XRBs (see, e.g., a review by McClintock \& Remillard 2003). In the
case of high-$z$ RQQs, the central 
black hole is expected to be supermassive, with a mass of up to $\sim 10^{10}\,
M_{\odot}$ (Rees 1984). Such a high mass is suggested by the high
luminosities of high-$z$ RQQs (of the order of $\sim10^{48}$ ergs~s$^{-1}$),
which should not exceed the Eddington luminosity of $L_{\rm Edd}
\approx 1.3 \times 10^{46} \left ( M / 10^8\, M_{\odot} \right )$~erg~s$^{-1}$. 

Some of the disk photons gain energy in the inverse Compton process by
scattering off energetic electrons in a hot, optically thin
plasma. This mechanism is generally accepted to explain hard
power--law--like spectral X-ray tails extending beyond 100 keV
(see Mushotzky, Done \& Pounds 1993 for a review; Done 2001).  The spectra
commonly show cut-offs beyond $\sim 100$ keV in both Seyfert galaxies
and XRB (in their hard/low states), suggesting a thermal energy
distribution of Comptonizing electrons (Gierli\'nski et al.\ 1997,
Zdziarski et al.\ 1998).  Observed X-ray spectra of XRBs (soft/high states)
may extend to energies beyond $500$ keV (e.g., Gierli\'{n}ski et al.\
1999) with no cut-off, implying a non-thermal energy distribution or,
possibly, a hybrid (i.e., a mixture of thermal and non-thermal; Coppi
1999) electron distribution.

Some of the Comptonized X-ray photons are likely to illuminate the
accretion disk, where they can be reprocessed and mostly 
thermalized, thus increasing the disk flux.  Since the disk flux
provides the seed photons for the inverse Compton processes, a
feedback loop between the cold disk and the hot plasma is created (Haardt
\& Maraschi 1991, 1993).

Some of the illuminating X-ray photons will be reflected by Compton
scattering off cold electrons and emerge back from the disk, with a
characteristic spectrum (Lightman \& White 1988; George \& Fabian
1991).  The spectrum is composed of (1) a broad continuum
(``reflection hump'') peaking at 20--30 keV, whose shape is determined by a 
competition between photo-absorption in the soft X-ray band ($E<10$
keV) and Klein-Nishina effects on the Compton scattering
cross-section in the hard X-ray band, (2) the Fe K$_{\alpha}$
fluorescence/recombination line at 6.4--6.97 keV (depending on the Fe
ionization state), and (3) the Fe-K$_{\alpha}$ edge at $\sim7.1$--10
keV.  The fluorescence features can be further influenced by
relativistic effects and/or the ionization state of the reflecting medium
(Fabian et al.\ 1989; {\.Z}ycki \& Czerny 1994, Nayakshin 2000).

An unresolved issue is that of the geometry of the accretion flow,
i.e., a geometrical configuration of the two
phases of accreting plasma (cold, optically thick; and hot, optically
thin). There are two general scenarios considered in the
modeling: (1) the plane-parallel geometry, in which a hot corona is located
atop the accretion disk, which is assumed to extend down to the
innermost stable orbit ($r_{\rm in} = 3\, R_{\rm S}$ for a
Schwarzschild black hole, where Schwarzschild radius $R_{\rm S} \equiv 2GM/c^2$), and (2)
the spherical geometry, which assumes that the cold disk is truncated
at a certain radius $r_{\rm tr} > r_{\rm in}$ and surrounds a hot
inner plasma flow within $r < r_{\rm tr}$ (Shapiro et al.\ 1976;
Poutanen et al.\ 1997; Esin, McClintock \& Narayan 1997;
R\'{o}\.{z}a\'{n}ska \& Czerny 2000). In the first 
geometry, the hot corona sandwiching the disk may be continuous (e.g.,
Haardt \& Maraschi 1991) or ``patchy'' i.e., composed of a number of
active regions driven by, e.g., magnetic activity (Galeev, Rosner \&
Vaiana 1979).  A variant of the spherical geometry (case 2 above) is
the spherical cloud model of Collin-Souffrin et al.\ (1996), in which the
cold plasma forms small clouds accreting roughly spherically onto the
central X-ray source.
 
In principle, one can distinguish between the two basic geometries by
studying the X-ray spectrum in detail (see Done 2001 for a review). One
argument is based on the amplitude of the reflection component. The
amplitude is defined as 
\be
\label{eq:refl}
R \equiv \Omega/2\pi,
\ee
where $\Omega$ is a
solid angle subtended by the reflector as seen from the source of
X-rays. For plane-parallel geometry, $R$ should yield values close to
unity (the disk covers $\approx 2\pi$ from the point of view of the
X-ray source), whereas for spherical geometry $R<1$ would be
expected. Furthermore, the amount of broadening of the reflection
features gives an estimate of the inner radius of the accretion disk
(e.g., Done et al.\ 2000).  For example, the Fe
K$\alpha$ line would be narrower if the disk were truncated far away
from $r_{\rm in}$.

Another argument in determining the geometry may be based on the X-ray
photon index, $\Gamma$. A continuous, plane-parallel corona produces
rather soft X-ray spectrum, $\Gamma>2$, even when the entire
gravitational energy is dissipated in the corona (Haardt \& Maraschi
1991, 1993).  This is because the underlying cold disk provides a
lower limit to the cooling flux. The spectra in the spherical geometry
are harder, $\Gamma < 2$, because the disk flux entering the central
sphere is strongly reduced, compared to that in the plane-parallel geometry
(e.g., Poutanen et al.\ 1997).

However, these diagnostics do not determine the geometry in a unique
way, since it is possible to modify the continuous plane-parallel
corona scenario in such a way that the predicted spectra are hard,
with $\Gamma<2$, and the amplitude of reflection is reduced, $R<1$.
This may be realized in a number of scenarios: (1) the hot plasma may
form a patchy corona (Stern et al.\ 1995); (2) the cold disk may
be covered by a hot ionized skin, decreasing the effectiveness of the
reprocessing/thermalization (Nayakshin, Kazanas
\& Kallman 2000); or (3) the hot plasma may be outflowing at relativistic speeds,
so that relativistic beaming makes the emission pattern significantly
anisotropic (Beloborodov 1999a). In particular, both the spherical
geometry scenario and models 2 and 3 above, reproduce the
correlation observed between the X-ray spectral index and amplitude of
reflection of both Seyfert galaxies and XRB (Zdziarski et al.\ 1999).

An additional model diagnostic, which helps to break the degeneracy,
is the ratio of observed optical and X-ray fluxes, the X-ray loudness,
$\alpha_{\rm ox}$. This parameter is of more use for AGNs than for
XRBs, since the disk emission can be more easily 
observed in AGNs than in XRBs. This is especially true for high-$z$
quasars, where the big blue bump may be redshifted into the
observable optical band. On the other hand, the quality of X-ray data
from these objects is usually not sufficient to accurately determine
the amplitude of the reprocessed component (see Siebert et al.\ 1996,
Lawson \& Turner 1997, Cappi et al.\ 1997, Reeves \& Turner 2000,
Mineo et al.\ 2000, and Page et al.\ 2003 for attempts to fit reflected
component and/or iron K$_{\alpha}$ line to the data of quasars).\\

The ``complex'' models described above were applied to Seyfert
galaxies and low redshift RQQs with high S/N (signal-to-noise
ratio) data (recently, e.g., by Chiang 2002; Chiang \&
Blaes 2003; Janiuk, Czerny \& Madejski 2001).  So far the quasars' spectral
energy distribution, 
especially that of high redshift quasars, has been modeled by the thermal
emission from an accretion disk (Sun \& Malkan 1989, Bechtold et al.\
1994a, 1994b, Tripp et al.\ 1994), with the addition of a Comptonizing medium 
explaining the X-rays (Band \& Malkan 1988, Fiore et al.\ 1995). There
have been no systematic studies of the parameter space of these models
appropriate for high$-z$ quasars, while the number of the observations
has been increasing.

In this paper we consider a geometry with a hot inner flow surrounded by a
truncated accretion disk covered with a hot corona. We apply the model to the
high redshift RQQs. Similar geometry was applied by Chiang 
(2002) and Chiang \& Blaes (2003) to Seyfert galaxies. Their
Compton scattering medium was located within the central sphere, which
was surrounded by an accretion disk. They
extended the model of Zdziarski et al.\ (1999) by introducing viscous
dissipation in the cold disk and modeled X-ray data 
simultaneously with the optical/UV data. In our model we include two
Comptonizing media, the sphere and the corona.
We explore the model parameter space and identify the model
parameters suitable to generate the observed spectra of RQQs.
The observational constraints include three quantities: the
X-ray photon index, $\Gamma$, the X-ray loudness, $\alpha_{\rm ox}$,
and the optical/UV luminosity, $l_{\rm UV} \equiv \log \left ( \nu
L_{\nu} \right )$ at 2500 \AA\, in the rest frame. Optical/UV luminosities
of high-$z$ RQQs are of the order of $l_{\rm UV} 
\gtrsim 46.0$--47.0 (B03 and references therein). B03 and Vignali et
al.\ (2003a, hereafter V03; 2003b) derived the mean $\sim 2$--30 keV
rest frame X-ray photon index to be in the range $\sim 1.5$--2. 
The X-ray loudness was found to be $\sim 1.7$--1.8 (B03; V03; Vignali et
al.\ 2003b).

The structure of the paper is as follows: In Section \ref{sec:model}, we
describe the geometry of accretion flow and describe the model.
In Section \ref{sec:results}, we present the results of computations, and discuss the
constraints imposed on the parameter space by the observations. In
Section \ref{sec:app}, we apply the spherical geometry accretion model to the
high-$z$ RQQ data of B03. Finally,
Sections \ref{sec:disc} and \ref{sec:concl} contain discussion and concluding remarks, respectively.

In a subsequent paper (Sobolewska et al.\ 2004, hereafter Paper II), we study the
accretion disk model covered by a non-uniform corona composed from hot
clouds (a patchy corona).\\ 

\section{Model}
\label{sec:model}

\begin{figure*}
\epsscale{0.8}
\plotone{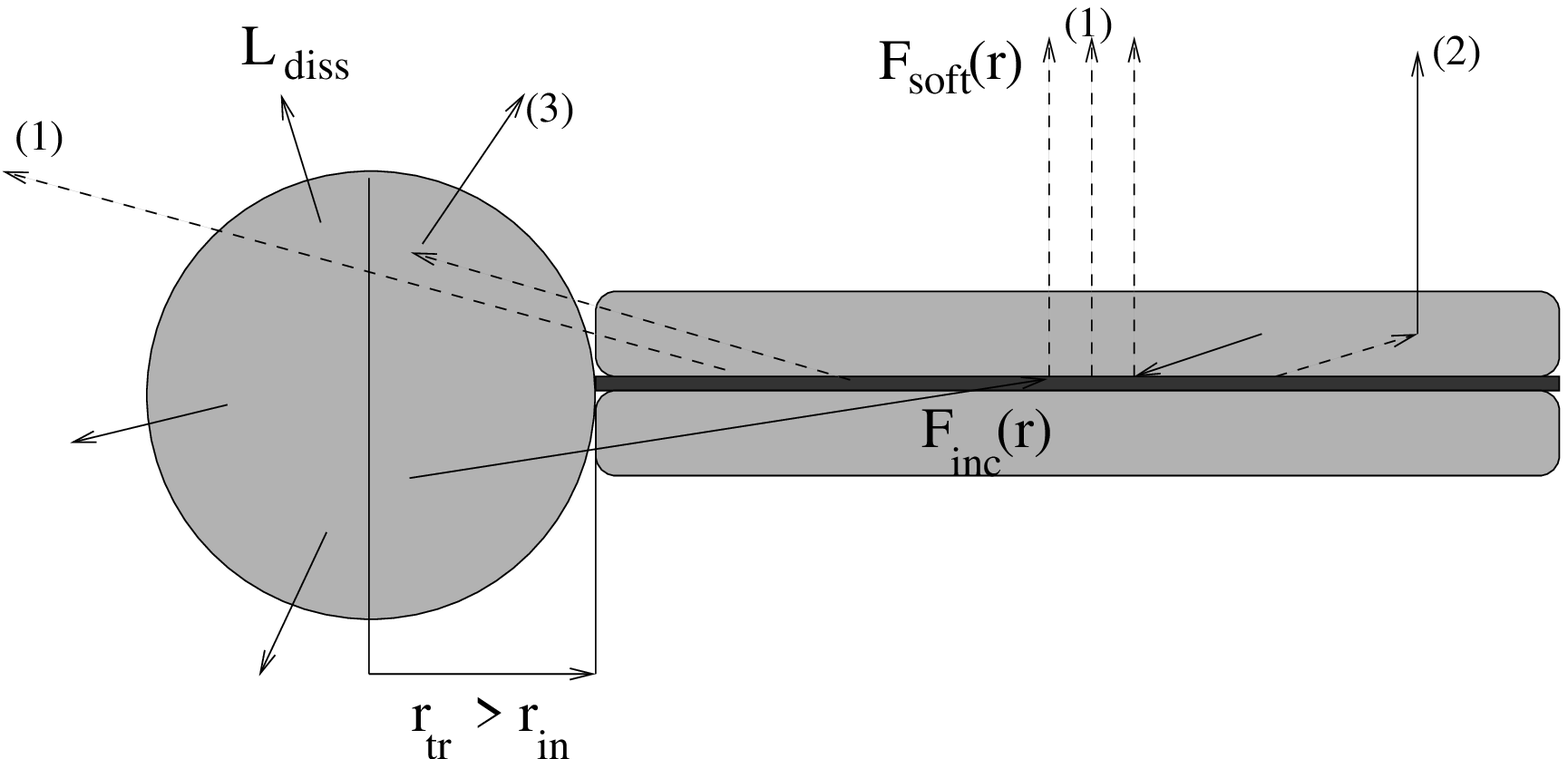}
\caption{Hot inner flow geometry. The accretion disk evaporates at
a radius $r_{\rm tr}>r_{\rm in}$ and forms a hot, semispherical inner flow. The
total luminosity of the hot inner flow is $L_{\rm diss}$. The 
truncated disk is covered with a hot corona at radii $r>r_{\rm tr}$. At
each radius $r>r_{\rm tr}$, three components contribute to the soft disk
flux $F_{\rm soft}(r)$: a viscous dissipation in the disk; reprocessing of
the hard coronal radiation; and reprocessing of the hard radiation from the
inner flow, $F_{\rm inc}(r)$. The spectrum is composed of (1) the 
soft disk blackbody-like radiation that escapes plasma without being
scattered, (2) the component Comptonized in the hot corona, and (3) the
component Comptonized in the hot inner flow.}
\label{fig:geom}
\end{figure*}

We consider an accretion flow consisting of (1) a cold, optically thick 
accretion
disk with (2) a hot corona above it. At a radius $r_{\rm tr}$ greater than the
radius of the innermost stable orbit, $r_{\rm in}$, the disk evaporates
completely to (3) a hot inner flow. The geometry is schematically
presented in Figure~\ref{fig:geom}. The computed spectra are a
superposition of the thermal disk radiation 
that escapes the hot plasma (both the corona and the inner flow) without
being scattered, and the Comptonized component. The spectra are
computed in a face-on view.

Throughout the paper we assume the Schwarzschild geometry, the
innermost stable circular orbit is 
located at $r_{\rm in} = 3\, R_{\rm S}$, where $R_{\rm S}=\frac{2GM}{c^2}$, and 
the accretion efficiency is $\epsilon$=1/12 (unless stated
otherwise). We define the Eddington accretion rate as 
$\epsilon\dot{M}_{\rm Edd} c^2 = 1.3 \times 10^{46}\frac{M}{10^8\,
  M_{\odot}}$~ergs~s$^{-1}$ and use the unitless quantity $\dot{m} =
\dot{M}/\dot{M}_{\rm Edd}$ to describe the accretion rate.\\

\subsection{The disk component}

At each radius $r>r_{\rm tr}$, we assume that the hot corona above the
accretion disk dissipates a fraction $f$ of the locally
released gravitational energy and that the remaining energy is dissipated in the 
disk.

The total dissipative luminosity available between the innermost
stable circular orbit, $r_{\rm in}$, and the transition radius, $r_{\rm tr}$, is
\be
L_{\rm available}=4\pi\int_{r_{\rm in}}^{r_{\rm tr}}F_{\rm visc}(r)rdr,
\label{lavailable}
\ee
where $F_{\rm visc}(r) = \frac{3GM\dot{M}}{8{\pi}r^3}\left
(1-\sqrt{\frac{r_{\rm in}}{r}} \right )$. We calculate the total
luminosity dissipated in the hot inner flow as
\be
L_{\rm diss} = \delta L_{\rm available},
\label{ldiss}
\ee
where $0 < \delta < 1$ is the radiation efficiency correction to the
Keplerian efficiency within the inner flow.

The cold disk ($r>r_{\rm tr}$) is illuminated by both the hot inner
flow and the hot corona. Part of this illuminating flux is
reprocessed in the disk. Therefore, at each radius three contributions
account for the total accretion disk soft flux: (1) viscous
dissipation in the disk, (2) reprocessing of the hard coronal
radiation, and (3) reprocessing of the hard radiation originating in
the inner flow (see Figure~\ref{fig:geom}). The total soft disk flux
is thus given by
\begin{eqnarray}
F_{\rm soft}(r) & = & (1-f)F_{\rm visc}(r)+f{\eta}(1-a)F_{\rm visc}(r)+ 
\nonumber \\
            & + & (1-a)F_{\rm inc}(r),
\label{fsoft}
\end{eqnarray}
where $a$ and $\eta$ are the energy-integrated disk albedo and the Compton
anisotropy parameter (Haardt \& Maraschi 1991), respectively (we 
assume typical values of $a=0.2$ and $\eta=0.5$; Haardt \& Maraschi 1993), 
while
$(1-a) F_{\rm inc}(r)$
is the flux from the inner flow intercepted by the disk annulus at $r$ and 
$F_{\rm inc}(r) = \frac{3 L_{\rm diss}}{16\pi^2 r_{\rm tr}^2 r_{\rm s}}
\int_{0}^{\alpha_{\rm max}} \sin\alpha \cos{\alpha}d\alpha
\int_{-\phi_{\rm max}}^{\phi_{\rm max}} l d\phi$ (Chiang 2002). For
completeness, we give expressions for $\alpha_{\rm max}$, $\phi_{\rm max}$,
and $l(r,\alpha,\phi,r_{\rm s})$ in the Appendix.
Derivation of $F_{\rm inc}$ was performed for a uniform, isotropically
radiating inner flow.\\ 

We assume that in the disk/corona region the accretion disk radiates
locally as a blackbody with a temperature $T=(F_{\rm
soft}/\sigma_{\rm SB})^{1/4}$, where $\sigma_{\rm SB}$ is the
Stefan-Boltzmann constant, and $F_{\rm soft}$ is given by equation
(\ref{fsoft}). This local disk emission is Comptonized in the
corona. The input spectrum for the Comptonization in the inner flow is
determined as a superposition of local blackbodies originating in the
disk at radii $r_{\rm i}$, between $r_{\rm tr}$ and $r_{\rm out}$, chosen
arbitrarily to be $10^5\, R_{\rm S}$. Each contributing blackbody is
normalized to the luminosity intercepted by 
the inner flow from a given radius, $L_{\rm soft}(r_{\rm i})$, given by
\be
L_{\rm soft}(r_{\rm i}) = 4\pi\int_{r_{\rm i}}^{r_{\rm
i+1}}g(r)\frac{F_{\rm soft}(r)}{\pi}rdr, 
\label{lsoftr}
\ee
where $g(r)$ is a geometrical factor (Chiang 2002; see also the
Appendix). These contributions sum to the total disk luminosity
intercepted by the hot inner flow, i.e.,
\be
L_{\rm soft} = \sum_{r_{\rm i}} L_{\rm soft}(r_i).
\label{lsoft}
\ee

\subsection{The Comptonized component}

In order to compute the Comptonized component, we use the empirical formulas
of Beloborodov (1999b) that 
relate the amplification factor, $A$, the photon index, $\Gamma$, and the
Compton parameter,
\be
\label{equ:ypar}
y=4\theta_e(1+4\theta_e)\tau(1+\tau)
\ee
(where $\theta_e=kT_e/m_ec^2$): 
\be
\label{b1}
\Gamma=\frac{7}{3}(A-1)^{-1/10},
\ee
\be
\label{b2}
\Gamma=\frac{9}{4}y^{-2/9}.
\ee

In general, the amplification factor is defined as follows:
\be
A=\frac{L_{\rm soft}+L_{\rm diss}}{L_{\rm soft}},
\label{Adef}
\ee
where $L_{\rm diss}$ stands for the total luminosity dissipated in the hot
plasma and $L_{\rm soft}$ denotes the soft luminosity that enters the hot
plasma from the accretion disk.

We perform our analysis for electron temperatures in the hot plasma of
$kT_e = 150$ and 500 keV.
We compute the optical depth of the plasma,
$\tau$, and the photon index, $\Gamma$, combining equations
(\ref{equ:ypar}), (\ref{b1}), and (\ref{b2}).

The relative strengths of the power-law-like high energy tail and the soft 
input spectrum
escaping the hot plasma without any shift in the energy are determined by
the probability, $P_{\tau}$,  that a soft photon is scattered to a higher
energy in the Comptonization process. To calculate P$_{\tau}$, we use the
formula 
\be
\Gamma-1 = -\frac{\ln P_{\tau}}{\ln {\Delta}E},
\label{alpha}
\ee
where $\Gamma$ is the photon index ($\Gamma-1$ is the spectral index) and
${\Delta}E=1+4\theta_e(1+4\theta_e)$ denotes the mean energy amplification
in a Thomson collision.

We derive formulae for the amplification factor, $A$, as follows. For the
disk/corona system $A$, defined by equation (\ref{Adef}), can be rewritten in
terms of the fraction of the energy released in the corona, $f$ (Haardt \&
Maraschi 1991): 
\be
A_{\rm corona}=1+\frac{f}{1-f[1-(1-a)\eta]}.
\label{Ad}
\ee

To obtain the amplification factor in the hot inner flow, $A_{\rm inner}$,
we substitute the $L_{\rm diss}$ specified by equation (\ref{ldiss})
and the $L_{\rm soft}$
given by equation (\ref{lsoft}) into the equation (\ref{Adef}).\\ 

\subsection{Efficient, hot, spherical flow}
\label{sec:eff}

In the simplest version, we assume that the hot inner flow efficiency is
comparable to the disk efficiency, i.e., $\delta = 1$ in equation
(\ref{ldiss}), and that the hot plasma forms a sphere with a radius $r_{\rm tr}$. 

The model parameters are\\
1. the mass of the black hole, $M$;\\
2. the accretion rate, $\dot{m}$;\\
3. the fraction of gravitational energy dissipated in the corona, $f$,
(assumed constant with radius);\\
4. the transition radius, $r_{\rm tr}$, at which the accretion disk
evaporates to the spherical inner flow; and\\
5. the plasma temperature, $kT_{\rm e}$ (same in both Compton components).\\

\subsection{Low efficiency, hot, semispherical flow}
\label{sec:loweff}

The hot plasma located within the truncation radius, $r_{\rm tr}$, may form
an advection dominated flow, which was found to be several times less
efficient than the accretion through the optically thick disk (see
review by Narayan, Mahadevan \& Quataert 1998).  Chiang (2002) inferred
the radiative efficiency in the Seyfert galaxy NGC 7469 to be $\sim
2\%$--3$\%$. Thus, we also consider a case with $\delta < 1$. In our
computations we assume $\delta=0.12$, which corresponds to
1\% accretion efficiency. In addition to the parameters
listed in Section \ref{sec:eff}, we have now also the accretion
efficiency correction, $\delta$. 

In both cases (efficient and low efficiency inner flow), we check the
possibility that the inner flow is flattened, i.e., the shape of the 
plasma within $r_{\rm tr}$ is ellipsoidal instead of
spheroidal. Flattened, disk-like distribution of Comptonizing plasma
was suggested by Chiang (2002) to explain spectra of the Seyfert galaxy
NGC 3516. This modification adds the semi-minor axis of ellipsoidal
flow, $r_{\rm s} < r_{\rm tr}$, to the list of model parameters.

The basic quantities that define the geometry in our approach are $L_{\rm
diss}$ and $L_{\rm soft}$. Allowing for $\delta < 1$ affects both $L_{\rm
diss}$ and $L_{\rm soft}$, since $L_{\rm soft}$ depends on $L_{\rm diss}$
through $F_{\rm soft}$; see equations (\ref{lsoftr}) and (\ref{fsoft}).
Considering a flattened shape of the inner flow affects only $L_{\rm soft}$,
i.e.\ the soft flux intercepted by the inner flow.\\

\section{Results}
\label{sec:results}

Based on the computed model spectra, we calculate spectral
characteristics, such as the X-ray loudness, $\alpha_{\rm ox}$, the
X-ray photon index, $\Gamma$, and the optical/UV luminosity, $l_{\rm
UV} \equiv \log \left ( \nu L_{\nu} \right )$ at 2500 \AA\, in the
rest frame. Then we explore the model parameter space to find
theoretical parameters compatible
with the observed values of $\alpha_{\rm ox}$, $\Gamma$, and $l_{\rm}$
in high-$z$ RQQs.

\begin{figure*}
\epsscale{0.8}
\plotone{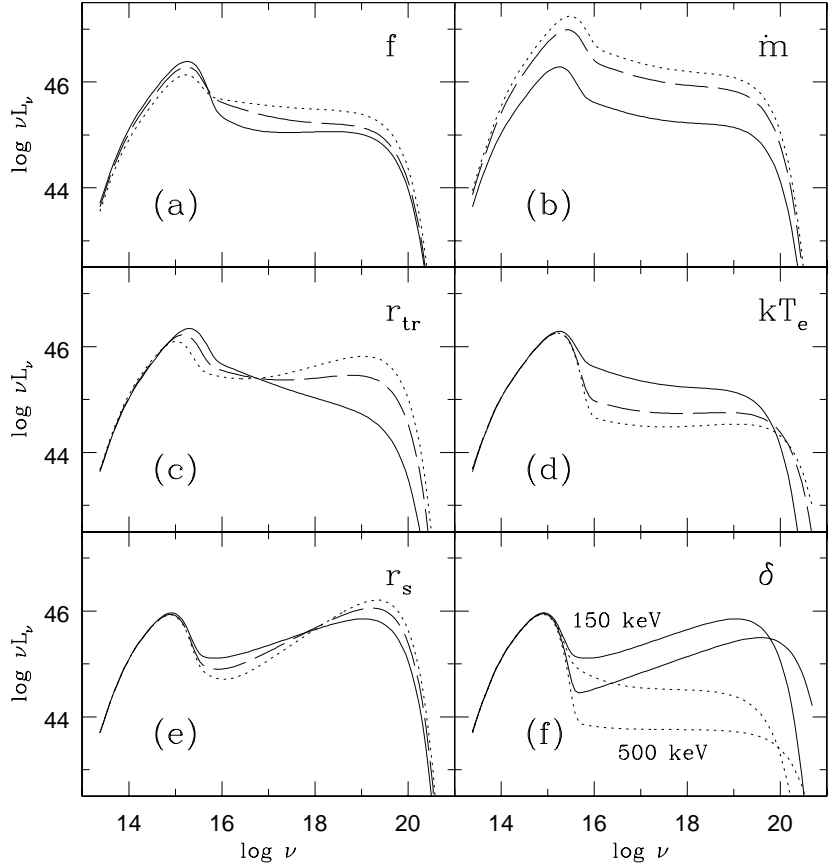}
\caption{Spectra computed in geometry with the hot inner flow. Unless stated
differently, the fixed parameters account for M=10$^{10}$ M$_{\odot}$, 
$\dot{m}$=0.1, $f$=0.5, $r_{\rm tr}$=7 R$_{\rm S}$, $r_{\rm s} = r_{\rm tr}$,
$kT_{\rm e} = 150$ keV, and $\delta = 1$. The panels show the dependence of 
the following model parameters: (a) the fraction of energy dissipated in
the clouds, $f$=0.1 ({\it solid curve}), 0.5 ({\it dashed curve}), and
0.9 ({\it dotted curve}); (b) the accretion 
rate, $\dot{m}$=0.1 ({\it solid curve}), 0.5 ({\it dashed curve}), and
0.9 ({\it dotted curve}); (c) the transition
radius, $r_{\rm tr}=5 R_{\rm S}$ ({\it solid curve}), $10 R_{\rm S}$ ({\it dashed curve}),
and $20 R_{\rm S}$ ({\it dotted curve}) R$_{\rm S}$; (d) the
electron temperature, kT$_e$=150 ({\it solid curve}), 350 ({\it dashed
  curve}), and 500 ({\it dotted curve}) keV; (e) the
semiminor axis of the ellipsoidal flow, $r_{\rm s}=25 R_{\rm S}$ ({\it
  solid curve}), $10 R_{\rm S}$ ({\it dashed curve}), $5 R_{\rm S}$
({\it dotted curve}), with $r_{\rm tr} = 25 R_{\rm S}$; (f) the
hot inner flow efficiency correction, $\delta = 1$ ({\it solid curves}), and $\delta <
1$ ({\it dotted curves}), for electron temperature of 150 ({\it upper
  curves}) and 500 keV ({\it lower curves}).} 
\label{fig:sphpar}
\end{figure*}

Figure~\ref{fig:sphpar} illustrates the sensitivity of the model
spectrum to
different model parameters. As expected, the bolometric luminosity
depends on the accretion rate, with almost no change in $\alpha_{\rm
ox}$ and $\Gamma$ (Fig.~\ref{fig:sphpar}b). The most dramatic spectral
change is due to variations in the truncation radius, $r_{\rm tr}$
(Fig.~\ref{fig:sphpar}c), the electron temperature (which physically
depends on the rate of cooling by soft photons;
Fig.~\ref{fig:sphpar}d), and the radiation efficiency of the central
sphere, $\delta$ (Fig.~\ref{fig:sphpar}f).

Below, we discuss in detail the results with respect to the data of
high-redshift RQQs. In particular, we search for values of the
theoretical model parameters that give $\Gamma < 2.3$ and $1.5
<\alpha_{\rm ox} < 1.8$.  Such a choice is motivated by the $3\sigma$
confidence intervals of $\Gamma$ given by V03
($1.7\leq\Gamma{\leq}2.3$), the results of Vignali et al.\ (2003b; $\Gamma 
= 1.86^{+0.41}_{-0.37}$), and the B03 sample, with a mean photon
index of $\Gamma = 1.50\pm 0.15$ (the error represents the 90\%
confidence interval). 
We choose the value of $\alpha_{\rm ox,min}
= 1.5$ for the lower limit, motivated by the value within the $3\sigma$
confidence interval found for RQQs in the Bright Quasar Survey (see Brandt et
al.\ 2002a and references therein). The upper limit of 
$\alpha_{\rm ox,max} = 1.8$ was chosen on the basis of the samples of B03, with 
$\alpha_{\rm ox}$=1.71$\pm$0.02 (the contributions to the error of
$\alpha_{\rm ox}$ 
come only from errors of the X-ray photon index and the normalization of the 
power law fit to the data; in addition, the 2 keV flux in
the source frame was calculated from the 1 keV flux in the observer frame,
assuming the photon indices found from fits, not the value of 2.2 as
assumed in B03), and V03, with $\alpha_{\rm ox}$=1.77$\pm$0.03 (we cite 
the value of $\alpha_{\rm ox}$ corrected by the authors
in a subsequent paper, Vignali et al.\ 2003b).\\ 

\subsection{Efficient, Hot, Spherical Flow}
\label{sec:reseff}

\begin{figure*}
\epsscale{0.3}
\plotone{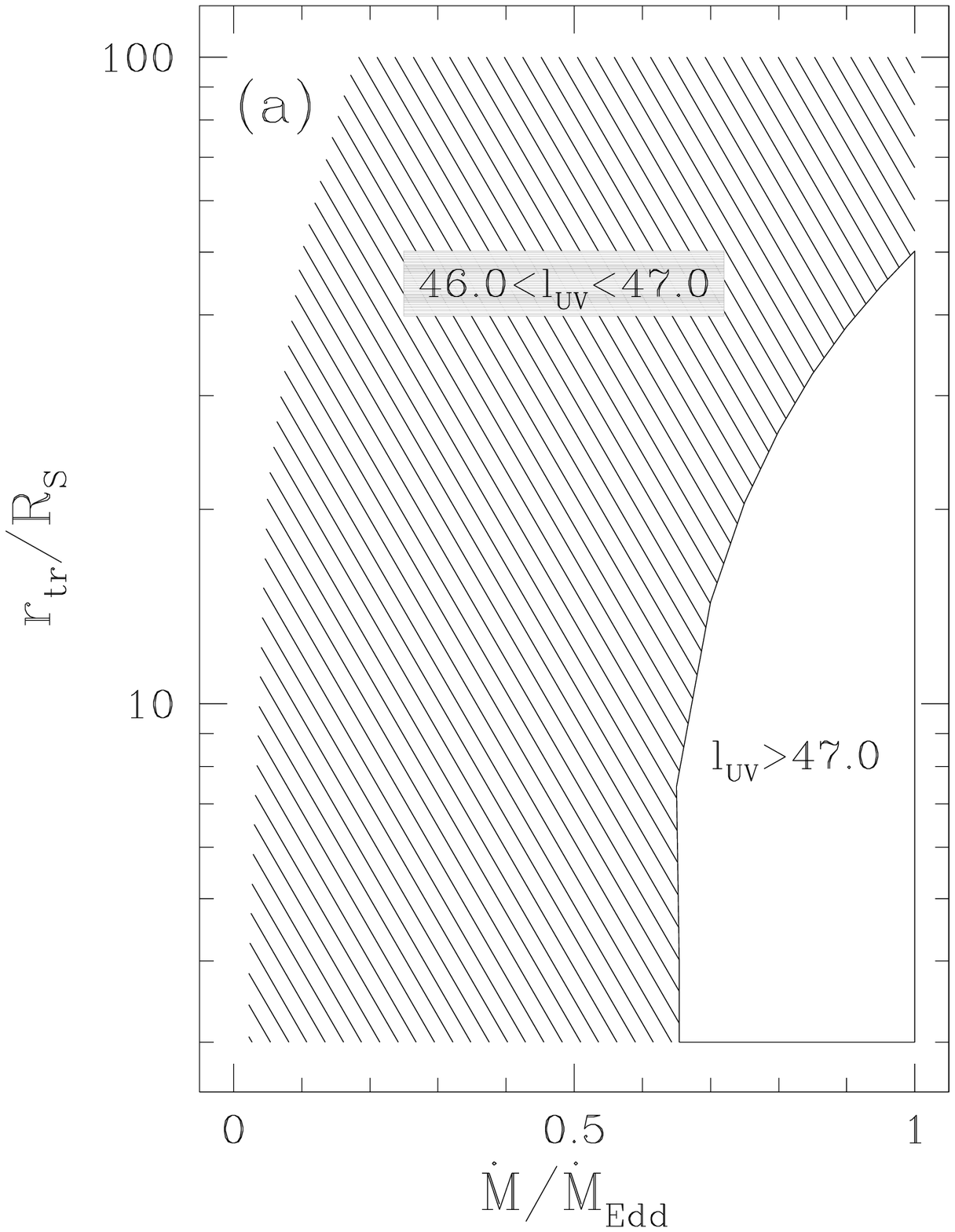}
\epsscale{0.3}
\plotone{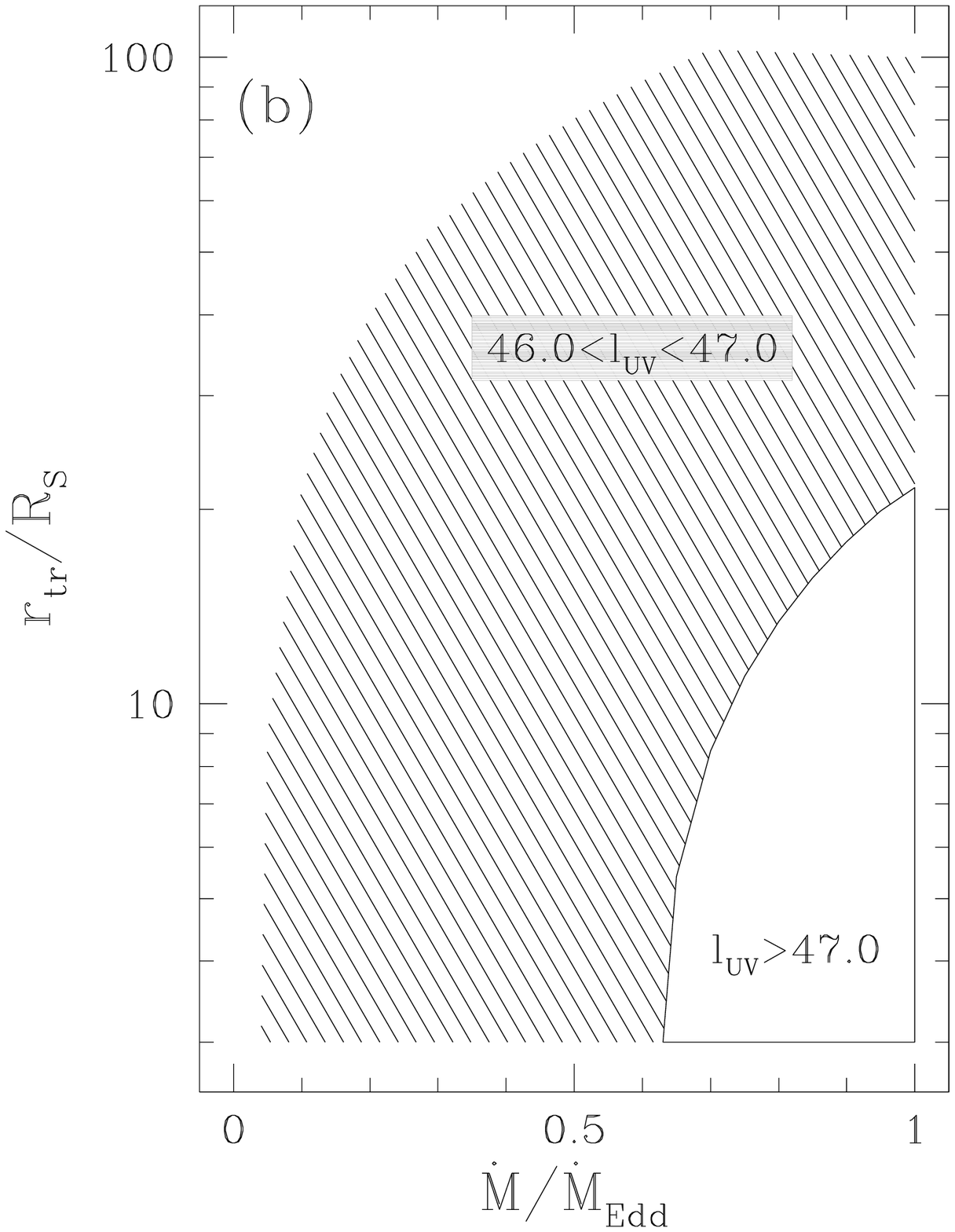}
\caption{Dependence of the optical/UV luminosity on the disk truncation radius, 
$r_{\rm tr}$, and the accretion rate, $\dot{m}$, for (a) $\delta = 1$ and (b)
$\delta < 1$ (low-efficiency flow). The shaded regions produce spectra with UV
luminosity 46.0$<l_{\rm UV}<$47.0. The solid curves enclose the part of parameter space with $l_{\rm UV}>47.0$. The
computations were done for a black hole mass of 10$^{10}$ M$_{\odot}$, the fraction of energy
released in the corona of $f = 0.1$, and the electron temperature of kT$_e$=150 keV. The result does not
depend on the electron temperature. If the corona becomes stronger ($f$ increases), the region with
$l_{\rm UV}>47.0$ disappears from the plot, and the most luminous objects cannot be fitted unless
M$>$10$^{10}$ M$_{\odot}$.} 
\label{fig:sphlum}
\end{figure*}

The rest frame ultraviolet luminosity, $l_{\rm UV} \equiv
\log\nu{L_{\nu}}$ at $\nu = c / \lambda_{\rm 2500\AA}\,$, depends on
all model parameters except the plasma temperature, $kT_e$. For a
black hole mass of the order of $10^{10}\, M_{\odot}$, the luminosity
$l_{\rm UV}\sim 46.0$--47.0 can be obtained for any considered
accretion rate if the corona above the truncated disk dissipates no
more than about 10\% of the gravitational potential energy, and the
remaining $\sim 90$\% is dissipated in the disk (see Figure
\ref{fig:sphlum}a). If the corona dissipates most of the energy, the
UV bump becomes less pronounced, and the most luminous objects (with
$l_{\rm UV} \gtrsim 47.0$) cannot be explained even with Eddington
accretion rates unless the black hole mass is higher than 10$^{10}$
M$_{\odot}$.

Neither the X-ray slope, $\Gamma$, nor the X-ray loudness,
$\alpha_{\rm ox}$, depends on the accretion rate for a given truncation
radius and fraction of energy dissipated in the corona (see Figure
\ref{fig:sphpar}b). The change in the accretion rate influences only
the normalization of the spectrum.\\

The X-ray photon index, $\Gamma$, depends on the strength of the
corona, $f$, and the disk truncation radius, $r_{\rm tr}$
(Fig.~\ref{fig:sphpar}a and Fig.~\ref{fig:sphpar}c). The strength of
the corona needs to be fixed at $\sim 10$\% in order to match the
observed optical/UV luminosity (as stated above), so the only
parameter that determines $\Gamma$ is the disk truncation radius,
$r_{\rm tr}$. If the truncation radius increases, the amplification
factor of the hot inner flow, $A_{\rm inner}$, increases, and a harder
X-ray spectrum is produced. For $4\, R_{\rm S} < r_{\rm tr} < 100\,
R_{\rm S}$, the model gives an X-ray photon index ranging from $\sim
2.3$ to $\sim 1.6$.

The X-ray loudness, $\alpha_{\rm ox}$, depends on the strength of the corona 
above the disk, $f$, the truncation radius, $r_{\rm tr}$, and the plasma
temperature, $kT_{\rm e}$. The strength of the corona, $f$, is constrained
by the optical/UV luminosity. An increase in the disk
truncation radius, $r_{\rm tr}$, causes the X-ray loudness to
decrease. This is because an accretion disk truncated farther 
away from the black hole produces a less pronounced optical/UV
bump, and $l_{\rm UV}$ decreases. Simultaneously, the X-ray spectrum
hardens (as described above), and a higher 2 keV flux is produced (see Figure
\ref{fig:sphpar}c). 

Variations of the electron temperature influence significantly 
the X-ray loudness, $\alpha_{\rm ox}$, but do not influence the X-ray slope, 
$\Gamma$ (Fig.\ref{fig:sphpar}d).
The maximum value of the X-ray loudness for a temperature of 150 keV
is about 1.7.
For a higher temperature, $k T_{\rm e} = 500$ keV, the maximum X-ray
loudness, $\alpha_{\rm ox, max}$, could be as high as $\sim 2.2$.

The lack of the dependence of the X-ray slope on the plasma
temperature follows from the fact that in the model $\Gamma$ is
determined by the amplification factor; i.e., the X-ray slope is only
geometry dependent. The sensitivity of the X-ray loudness to the
plasma temperature can be explained by the fact that in thermal
plasmas spectral characteristics, such as the X-ray photon index, the
optical depth, and the electron temperature, are not independent. An
increase in the electron temperature is compensated for by a decrease in
the optical depth,
resulting in approximately the same X-ray slope. The higher electron
temperature implies the higher energy of the spectral cut-off.
The total energy must, however, be conserved. Therefore, the
normalization of the Comptonized component drops, resulting in the
X-ray quieter spectra. Plasma temperature variations only weakly
influence $l_{\rm UV}$, through the dependence on the amount of the
soft radiation that escapes the plasma without being Comptonized on
$kT_{\rm e}$ (see Equation (\ref{alpha})). However, this effect is not
significant enough to influence the results. 

\begin{figure*}
\epsscale{0.3}
\plotone{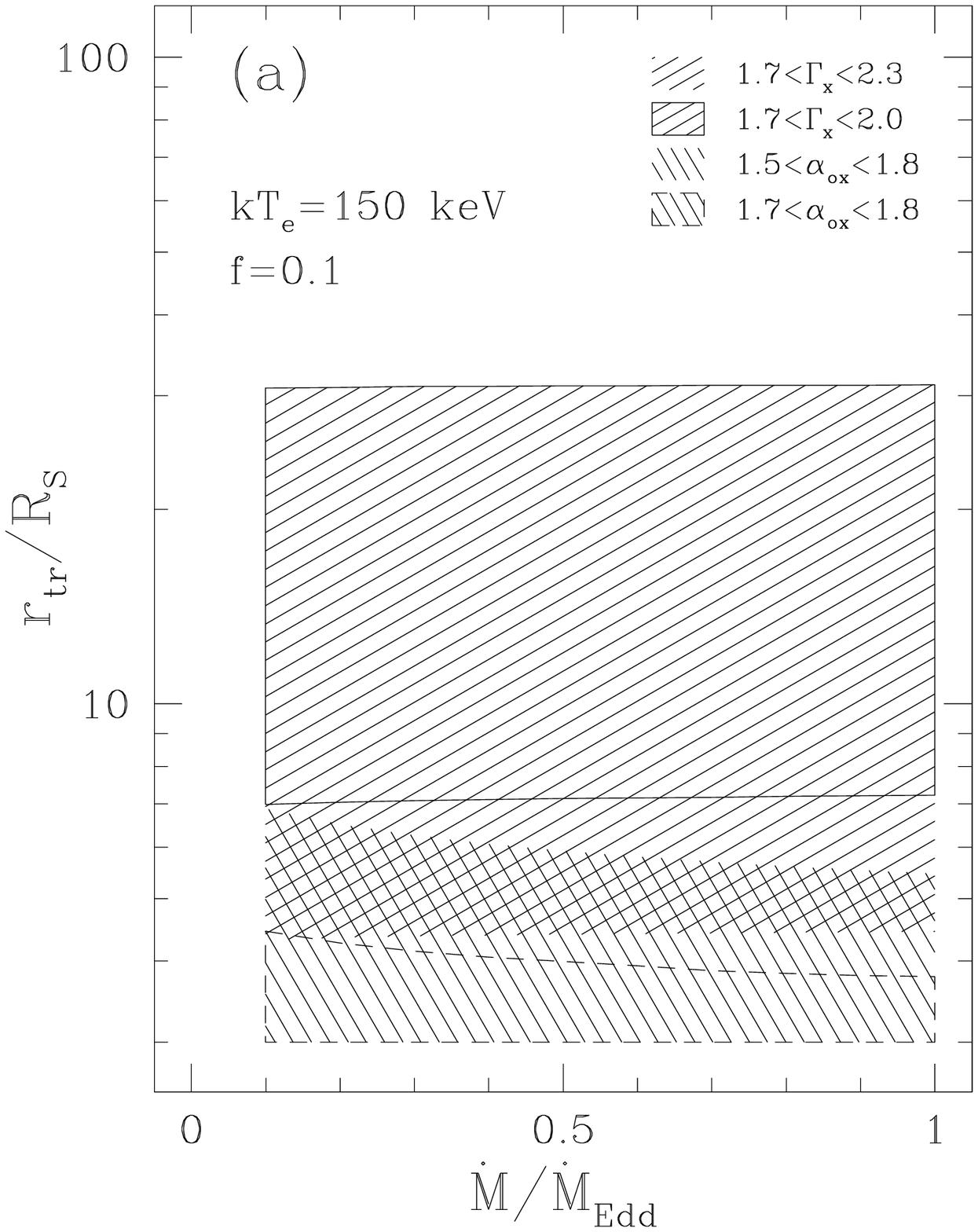}
\epsscale{0.3}
\plotone{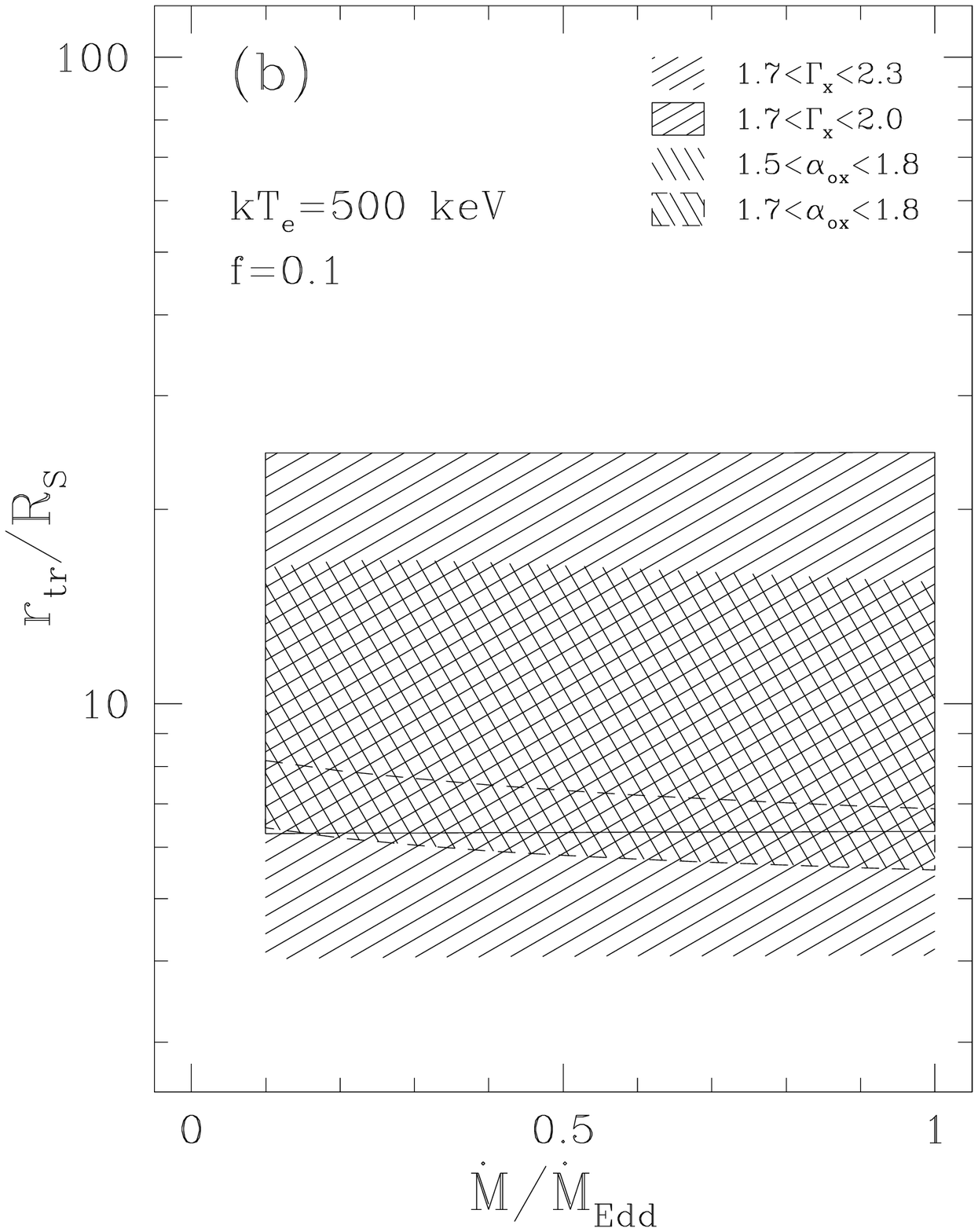}
\epsscale{0.3}
\plotone{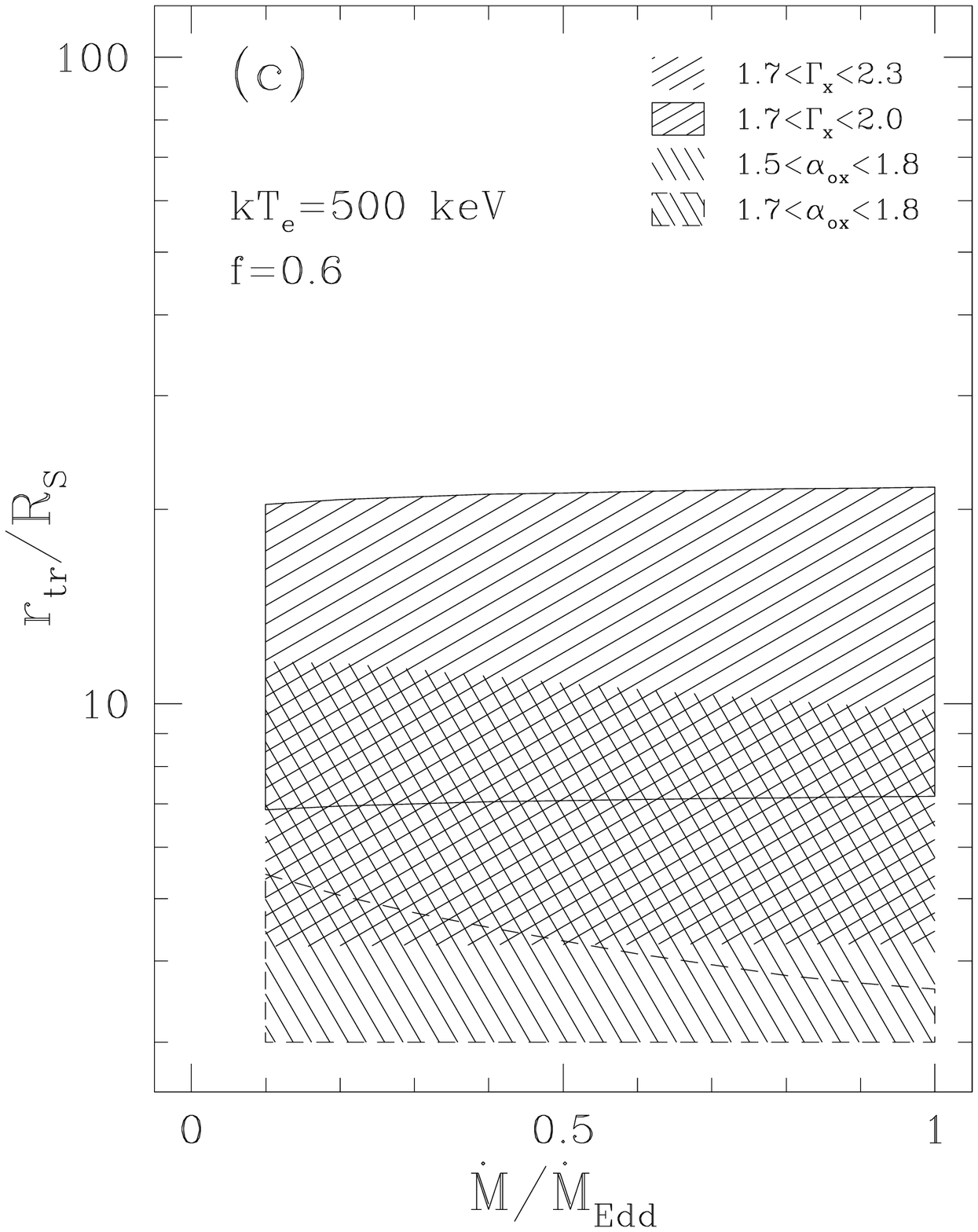}
\caption{Results for geometry with efficient, hot, spherical inner flow
($\delta = 1$, $r_{\rm s} = r_{\rm tr}$; Section \ref{sec:eff}),
  showing the overlapping of regions yielding $1.5 < \alpha_{\rm ox} < 1.8$ and
$1.7 < \Gamma < 2.3$. The solid curve encloses the part of the parameter space
with the hardest spectra ($\Gamma < 2$). The dashed curve encloses the region
with the X-ray quietest spectra ($\alpha_{\rm ox} > 1.7$).  The  
contours are calculated for $kT_{\rm e} = 150$ keV (a) and 500 keV (b
and c), and a fraction of energy dissipated in the corona of $f = 0.1$
(a and b), and $f = 0.6$ (c). The overlapping exists for $r_{\rm tr}
\leq 15\, R_{\rm S}$. For electron temperatures $\leq 150$ keV, the model can 
produce relatively soft and X-ray loud spectra with $\Gamma \sim 2.0-2.3$,
and $\alpha_{\rm ox} \sim 1.5-1.7$ (a). For higher electron temperatures,
harder ($\Gamma < 2.0$) and X-ray quieter ($\alpha_{\rm ox} \sim 1.7-1.8$)
spectra can be obtained (b). For a stronger corona and high electron
temperature, the model gives either hard or X-ray quiet spectra (c). The
computations were done for a black hole mass of $10^{10}\, M_{\odot}$.}
\label{fig:sphd1}
\end{figure*}

Figure \ref{fig:sphd1} summarizes the above results and shows two
observationally interesting regions, $1.5 \leq
\alpha_{\rm ox} \leq 1.8$ and $1.7 \leq \Gamma \leq 2.3$,
in the {\it truncation radius--accretion rate} plane.
The computations were performed for two electron temperatures,
$kT_{\rm e}=150$ and 500 keV, and two values of the fraction of
energy dissipated in the corona, $f=0.1$ and $f=0.6$. For an electron
temperature of 150 keV and $f=0.1$,
the two regions overlap if the transition radius is of the order of
several $R_{\rm S}$. The spectra are relatively
soft and X-ray loud, with $\Gamma \sim 2.0$--2.3 and $\alpha_{\rm ox}
\sim 1.5-1.7$ (Figure
\ref{fig:sphd1}a). For higher electron temperatures, the overlapping
region exists at higher truncation radii ($r_{\rm tr} \sim 7 -
15\, R_{\rm S}$ for $kT_{\rm e}=500$ keV). In addition, harder ($\Gamma <
2.0$) and X-ray quieter ($\alpha_{\rm ox}\sim$1.7-1.8) spectra can be
obtained (Figure
\ref{fig:sphd1}b). A stronger corona ($f=0.6$) requires a higher electron
temperature ($kT_{\rm e} \gtrsim 200$ keV) to produce spectra with $\Gamma 
\sim 2.0 - 2.3$ and $\alpha_{\rm ox} \sim 1.5 - 1.7$. For $kT_{\rm e} \gtrsim
 400$ keV, the model additionally gives spectra with $\alpha_{\rm ox} \sim
1.7 - 1.8$ and $\Gamma \sim 2.0 - 2.3$, or $\alpha_{\rm ox} \sim 1.5 - 1.7$
and $\Gamma \sim 1.7 - 2.0$ (Figure \ref{fig:sphd1}c). However, in case of
$f = 0.6$, the optical/UV luminosity is $l_{\rm UV} \leq 46.2$.\\

\subsection{Low-Efficiency, Hot, Semispherical Flow}

We compute $\alpha_{\rm ox}$, $\Gamma$, and $l_{\rm UV}$ in the low
efficiency hot inner flow model (Section \ref{sec:loweff}).
All the spectral trends with the strength of the corona, the accretion
rate, the truncation radius, and the electron temperature are
qualitatively the same as for the high efficiency hot flow. As in
Section \ref{sec:reseff}, a weak corona above the disk is required in order to fit
the optical/UV luminosity. The dependence of $l_{\rm UV}$ on the
truncation radius and the accretion rate for $f=0.1$ is shown in
Figure \ref{fig:sphlum}b.

Figure \ref{fig:sphpar}f shows dramatic changes in the X-ray slope,
$\Gamma$, and X-ray loudness, $\alpha_{\rm ox}$, with
reduced efficiency of the central hot flow. For a given truncation
radius, $r_{\rm tr}$, less heat is being supplied to the electrons
(some of the energy is advected); i.e., the total luminosity of the
hot flow decreases, $L_{\rm diss}(\delta < 1) < L_{\rm diss}(\delta =
1)$. The soft disk 
luminosity intercepted by the hot flow, $L_{\rm soft}$, also decreases,
because it depends on the hot flow luminosity through $F_{\rm inc}$
(see equations (\ref{lsoftr}) and (\ref{fsoft})). However, $L_{\rm
diss}$ decreases faster than $L_{\rm soft}$, since the latter
contains contribution from the viscous dissipation in the outer disk.
Thus, lowering the hot flow efficiency causes a decrease in the
amplification factor of the hot flow, $A_{\rm inner}$, and
leads to a softer X-ray spectrum. In addition, decreasing $L_{\rm
soft}$
lowers the normalization of the soft input spectrum for Comptonization
in the hot flow, which
results in a lower normalization of the X-ray continuum and X-ray
quieter spectra. This can be clearly seen in Figure
\ref{fig:sphpar}f for the plasma temperature of 500 keV.

\begin{figure*}
\epsscale{0.3}
\plotone{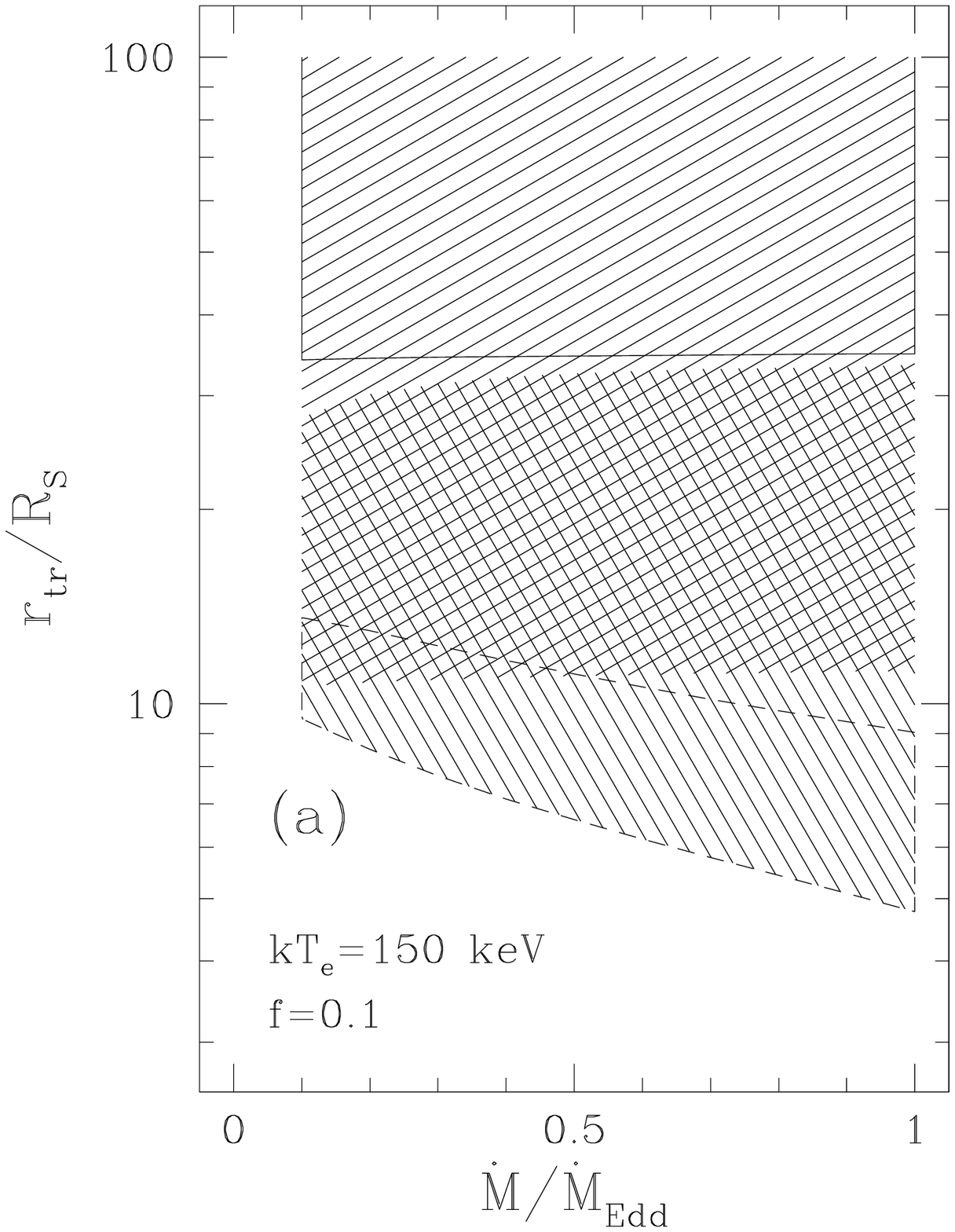}
\epsscale{0.3}
\plotone{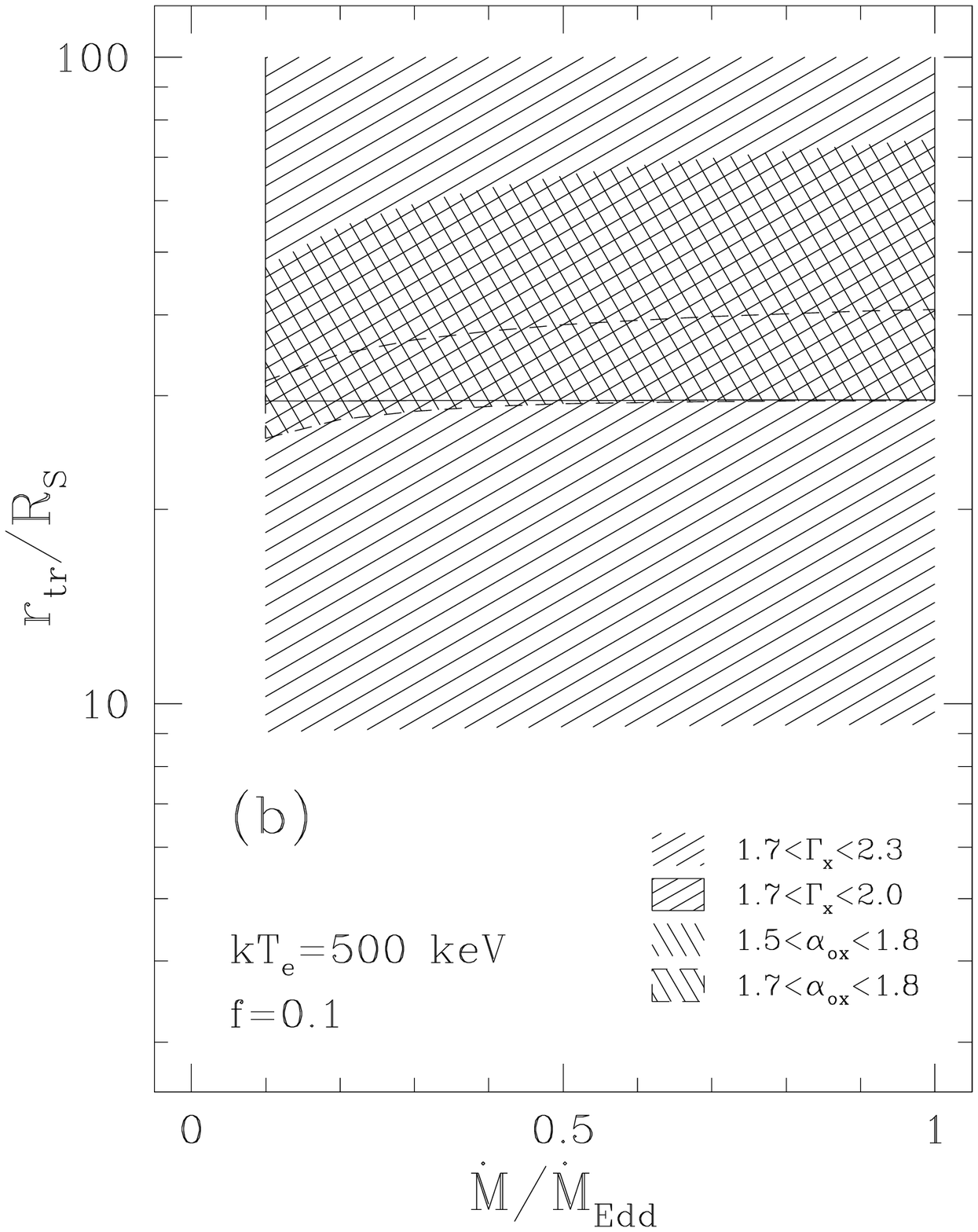}
\epsscale{0.3}
\plotone{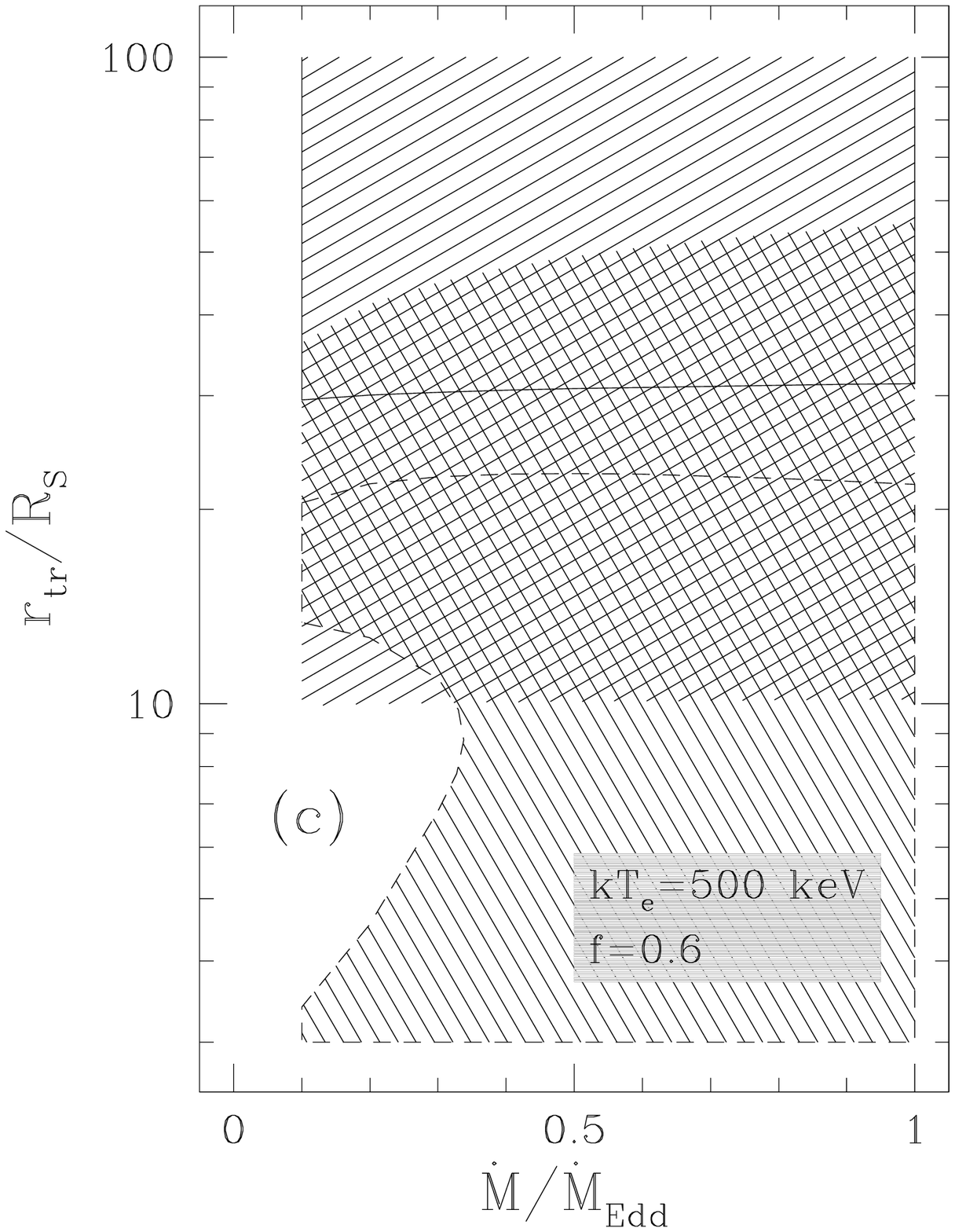}
\caption{Results for geometry with low-efficiency, hot, spherical inner
flow ($\delta < 1$, $r_{\rm s} = r_{\rm tr}$; Section
\ref{sec:loweff}), showing the overlapping of regions with 
$1.5 < \alpha_{\rm ox} < 1.8$ and $1.7 < \Gamma < 2.3$. The overall trends
are similar to those illustrated in Figure 
\ref{fig:sphd1}. However, the truncation radius is now $\sim 10 - 70\, R_{\rm S}$,
depending on the other parameters. The hardest and X-ray quietest spectra
are produced for $r_{\rm tr} \sim 30 - 40\, R_{\rm S}$ and high electron
temperatures (b). The
computations were done for a black hole mass of $10^{10}\, M_{\odot}$.}
\label{fig:sphd012}
\end{figure*}

To summarize, for a given truncation radius, $r_{\rm tr}$, 
lowering the efficiency of the hot inner flow leads to a softer X-ray
slope and X-ray quieter spectra. This result is illustrated in
Figure
\ref{fig:sphd012}. To produce spectra with the same $\Gamma$ and
$\alpha_{\rm ox}$ as in the case of 
the efficient flow (compare Figure
\ref{fig:sphd1}), a sphere with a larger radius is required. As in
Section \ref{sec:reseff}, a plasma temperature higher than $\sim 150$
keV and a fraction
of energy dissipated in the corona smaller than 
$\sim 10$\%
are implied by the data. However, the
transition radius now falls into the range of $\sim 10 - 70\, R_{\rm
S}$, depending on the strength of the corona, the plasma temperature, and
the accretion rate. The hardest, $\Gamma \sim 1.7 - 2.0$, and X-ray
quietest, $\alpha_{\rm ox} \sim 1.7 - 1.8$, spectra can be obtain for
higher electron temperatures. Figure
\ref{fig:sphd012}b shows the case of  $kT_{\rm e} = 500$ keV with the 
solution for $r_{\rm tr} \sim 30 - 40\, R_{\rm S}$. 

In the second modification to the efficient flow, we allow for a
flattening of the central sphere, e.g., an ellipsoidal flow. The
amplification factor now depends strongly not only on $r_{\rm tr}$,
but also on the semi-minor axis of the ellipsoidal flow, $r_{\rm s}$,
and on the ratio $r_{\rm s} / r_{\rm tr}$. For a given transition
radius there exists the following trend: the smaller the $r_{\rm s}$, the
harder the spectra (see Figure
\ref{fig:sphpar}e). This is because the luminosity, $L_{\rm soft}$,
drops, resulting in an increase of the hot flow amplification factor,
$A_{\rm inner}$ ($L_{\rm diss}$ remains constant). This geometry may
produce spectra as hard as $\Gamma=1.5$, but with $\alpha_{\rm ox}$
significantly smaller than that observed in high-$z$ RQQs if the hot inner
flow is efficient, $\delta=1$ (Section~\ref{sec:eff}).  If the
radiative efficiency is reduced, $\delta<1$
(Section~\ref{sec:loweff}), allowing for ellipsoidal central flow 
results in a shift of the regions indicated in Figure
\ref{fig:sphd012} toward slightly lower transition radii (i.e., it is
an intermediate case between efficient spherical inner flow and
efficient flattened inner flow). Thus, we
consider $r_{\rm tr} \sim 40\, R_{\rm S}$ to be an upper limit of the
truncation radius in the model that allows us to obtain the X-ray quietest
spectra with $\alpha_{\rm ox} \sim 1.7 - 1.8$.

We note that the objects with very hard photon indices, $\Gamma <
1.5$, cannot be explain by the model, because the X-ray loudness of
such spectra is always lower than $\alpha_{\rm ox} = 1.5$--1.8.\\

\section{Application to the data}
\label{sec:app}

\begin{figure*} 
\epsscale{0.45} 
\plotone{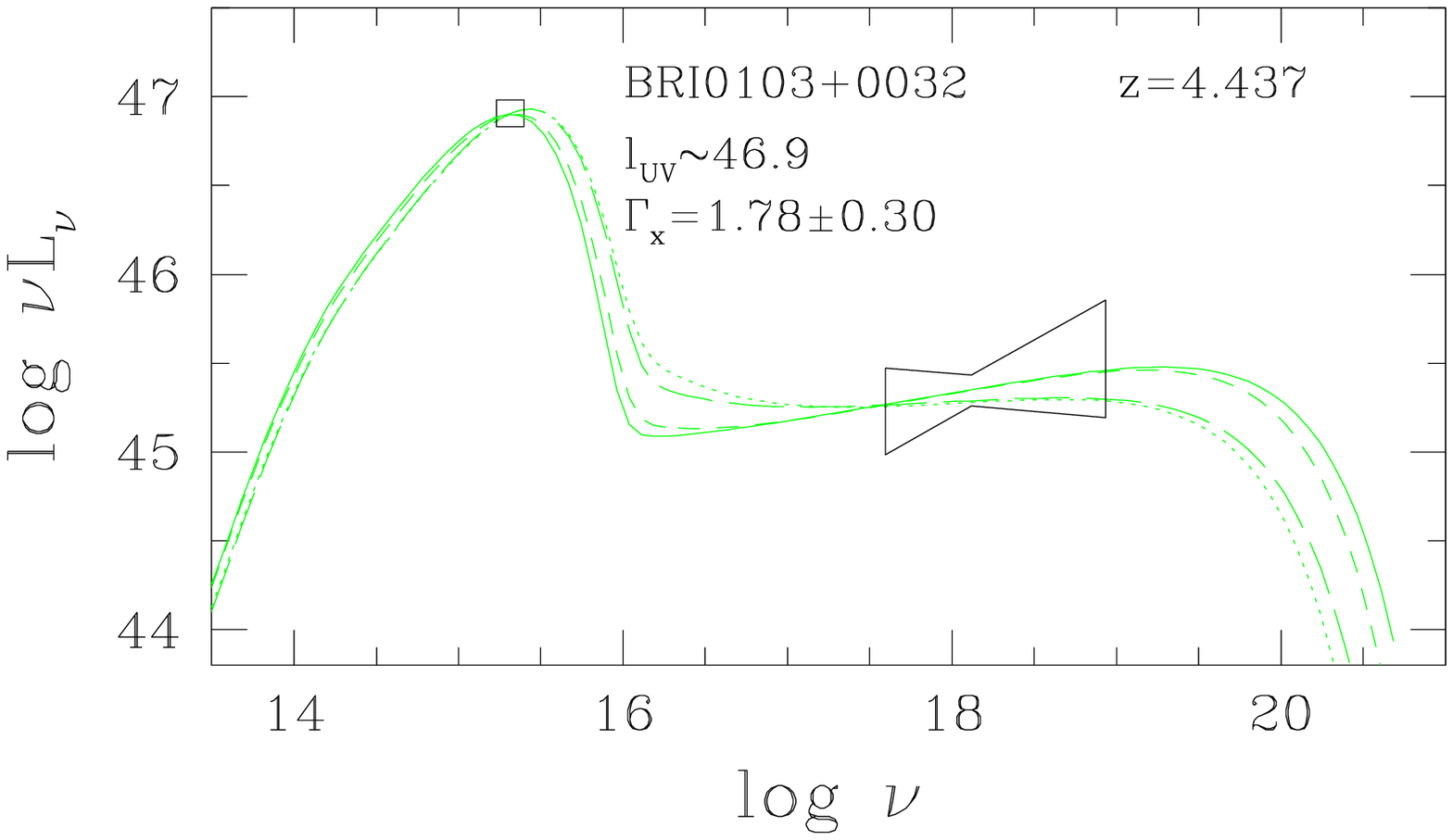} 
\epsscale{0.45} 
\plotone{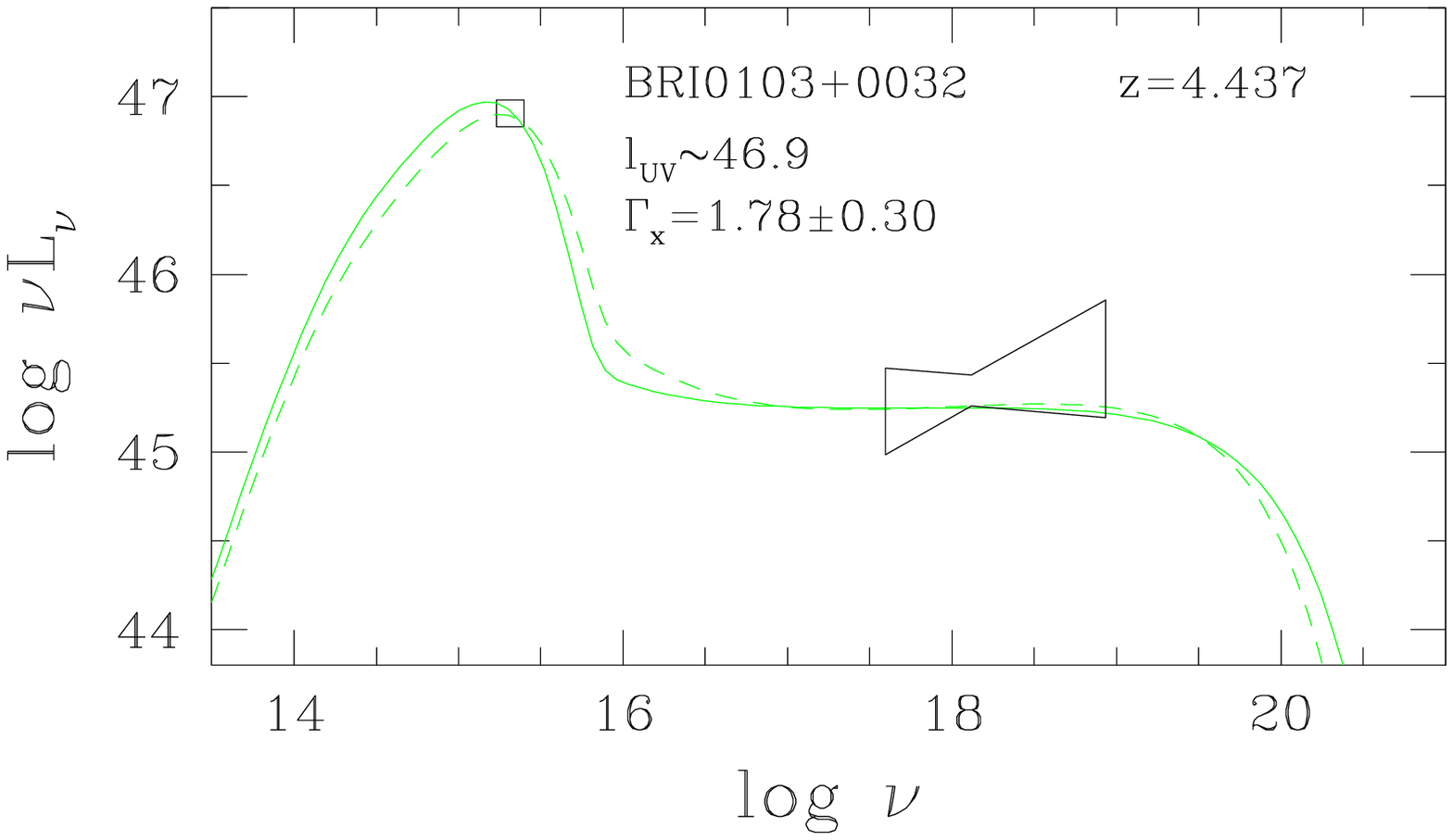}
\epsscale{0.45}
\plotone{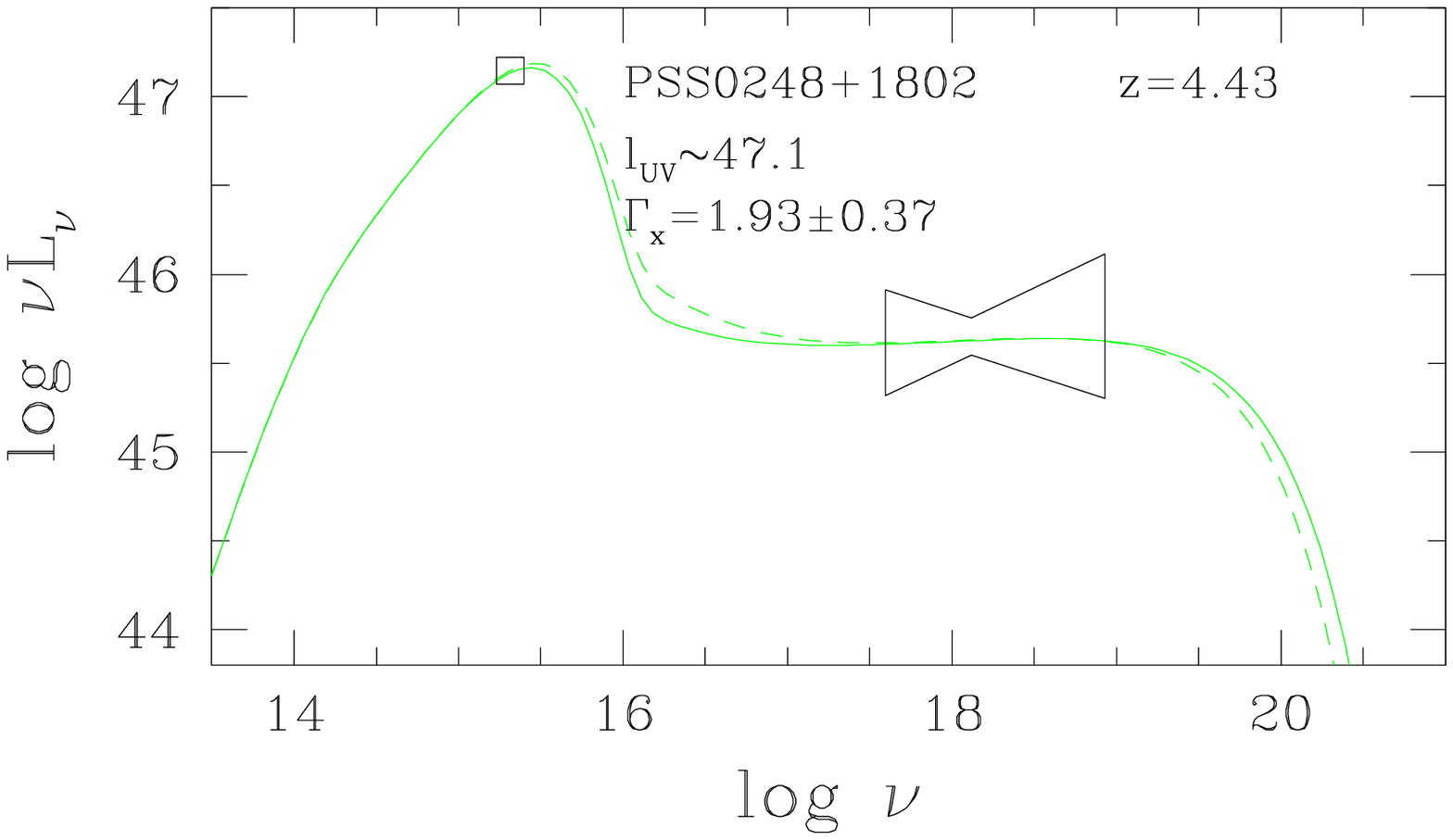}
\epsscale{0.45}
\plotone{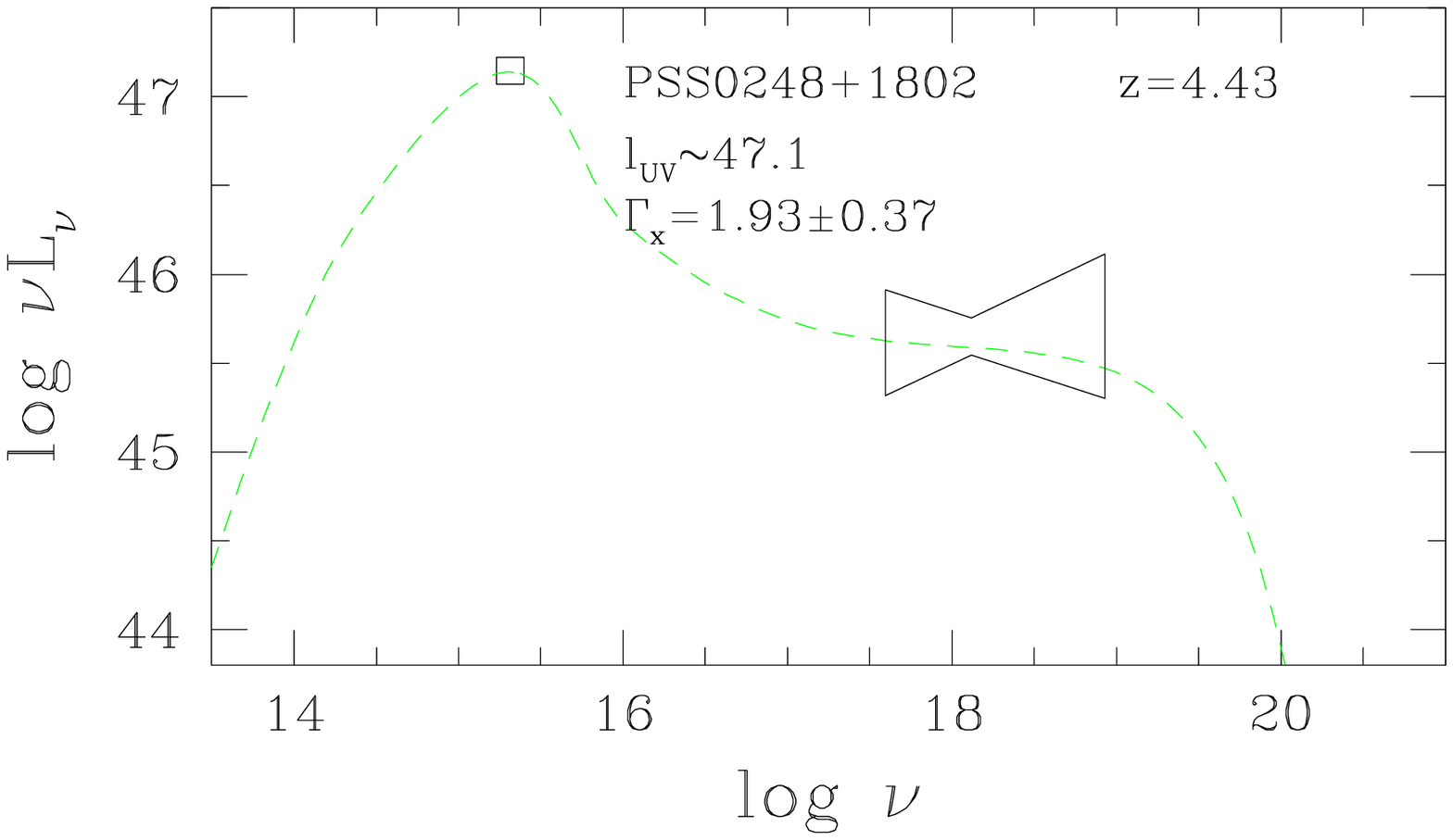}
\epsscale{0.45}
\plotone{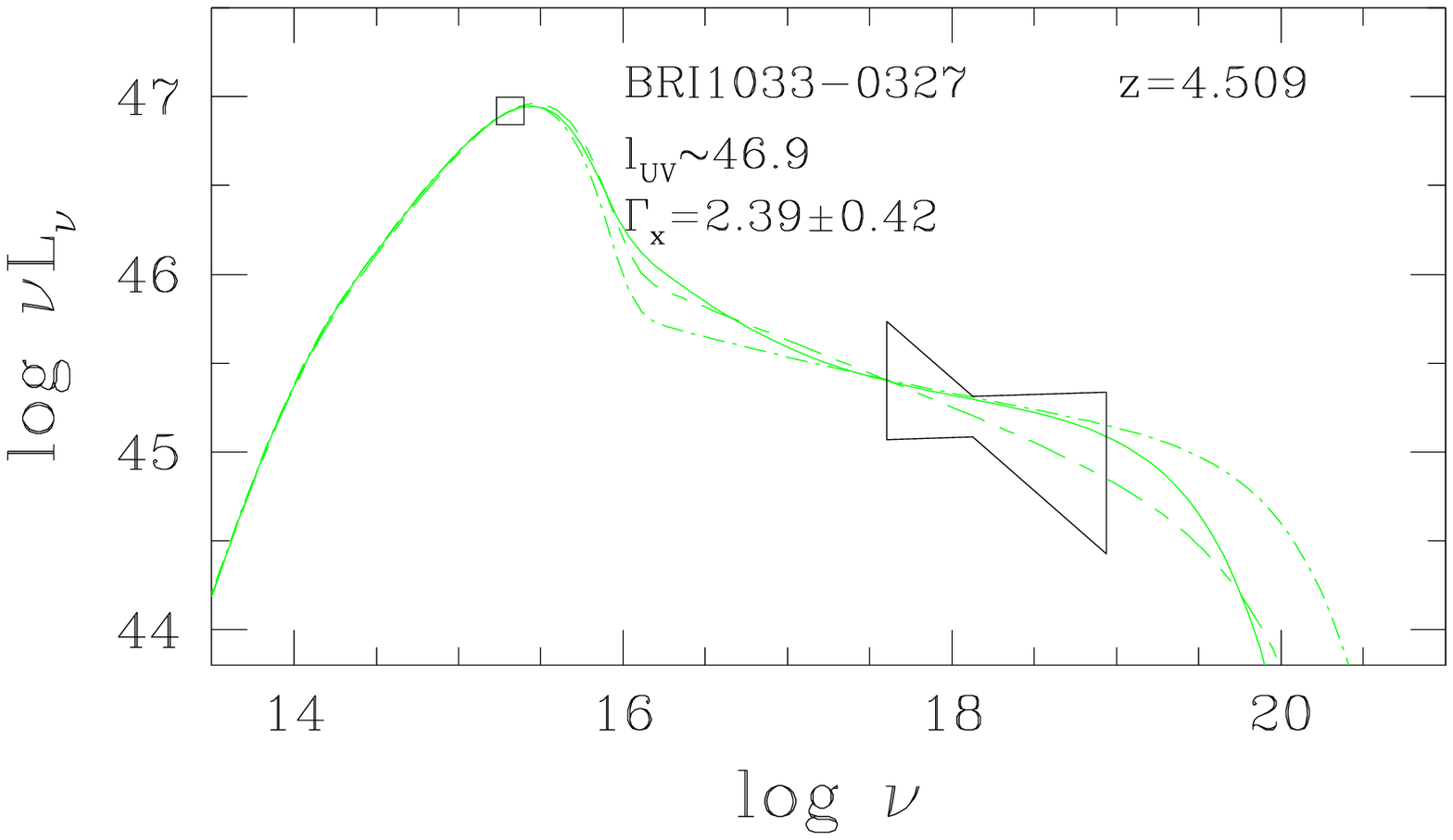}
\epsscale{0.45}
\plotone{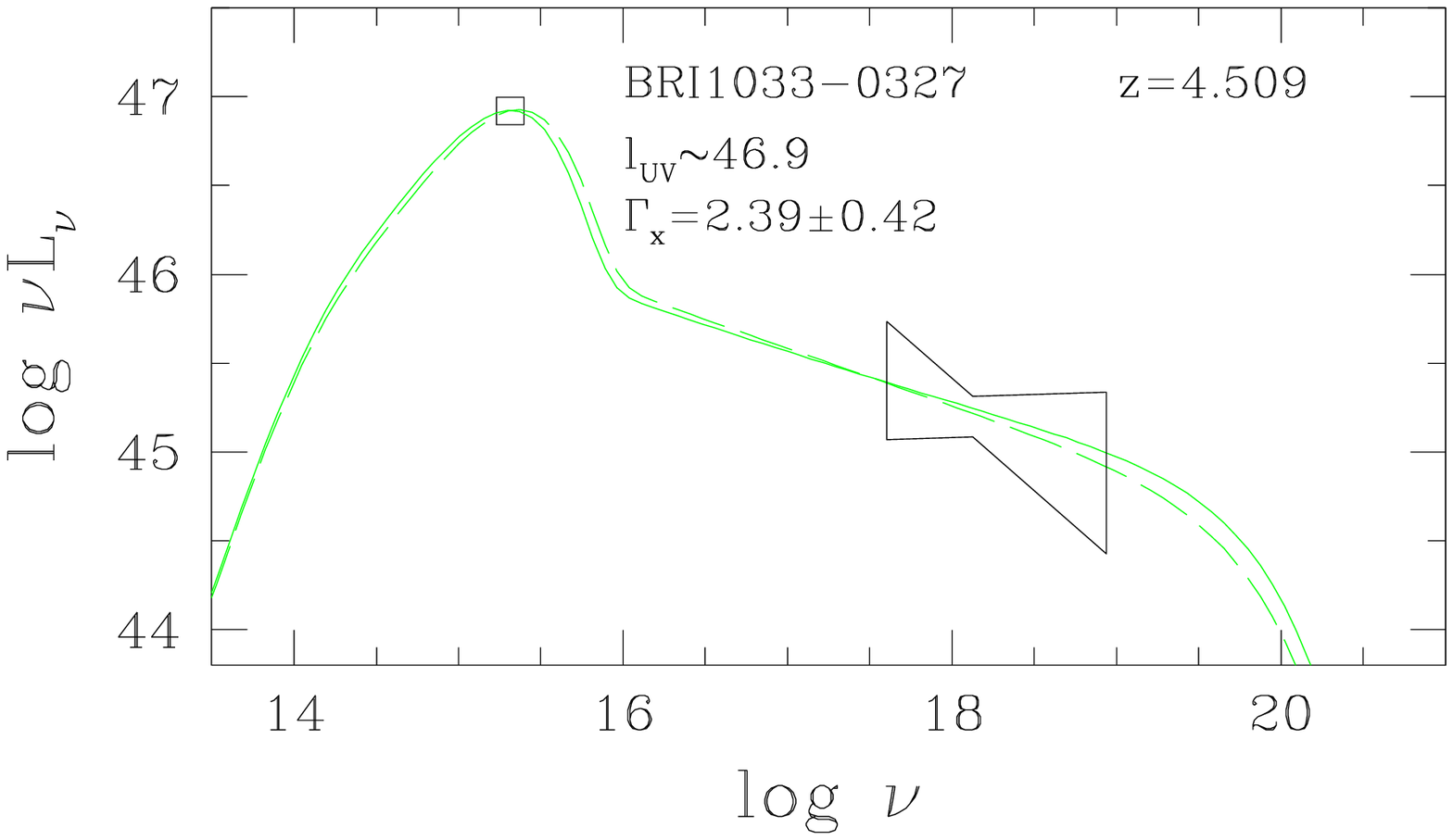}
\epsscale{0.45}
\plotone{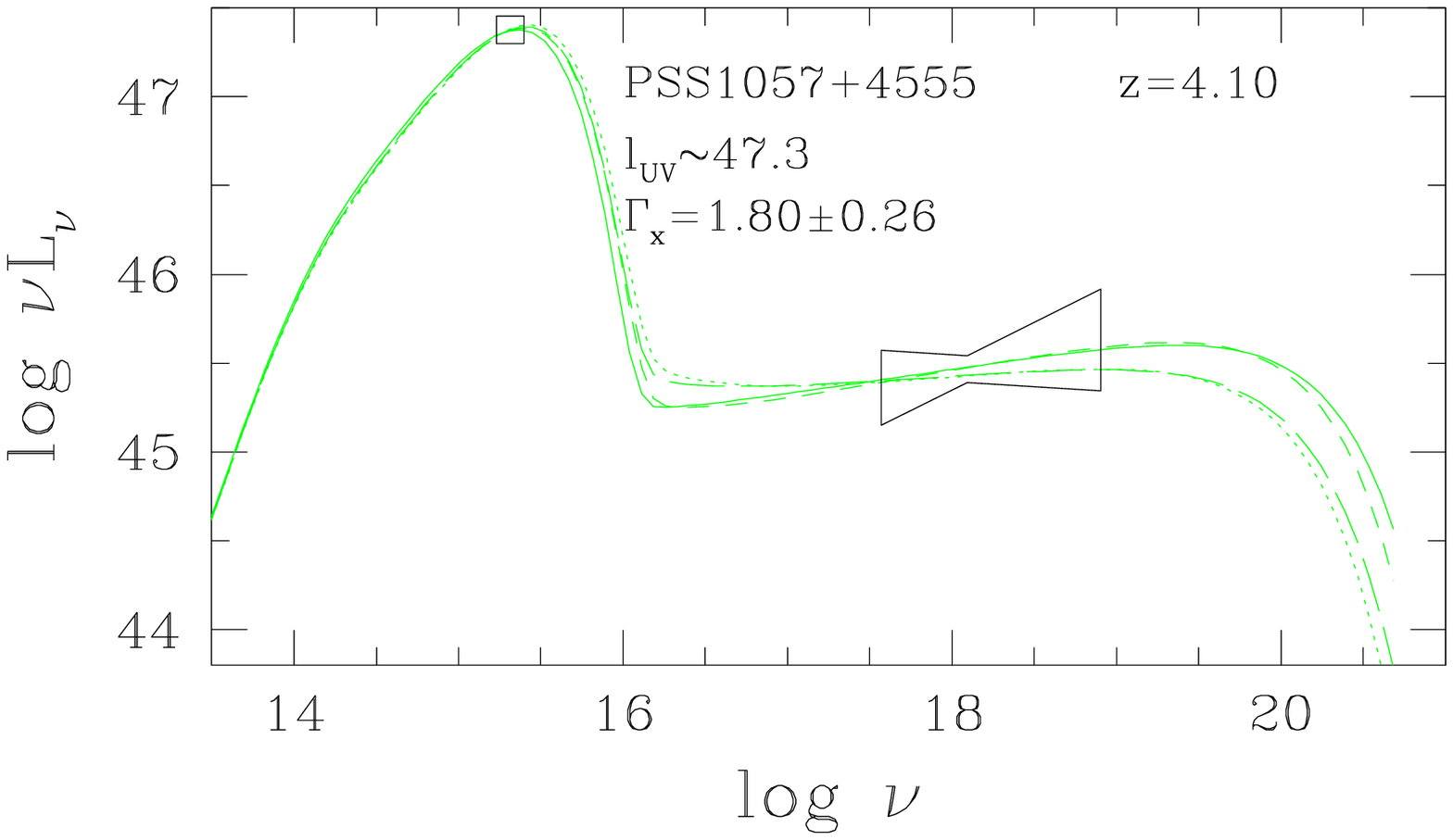}
\epsscale{0.45}
\plotone{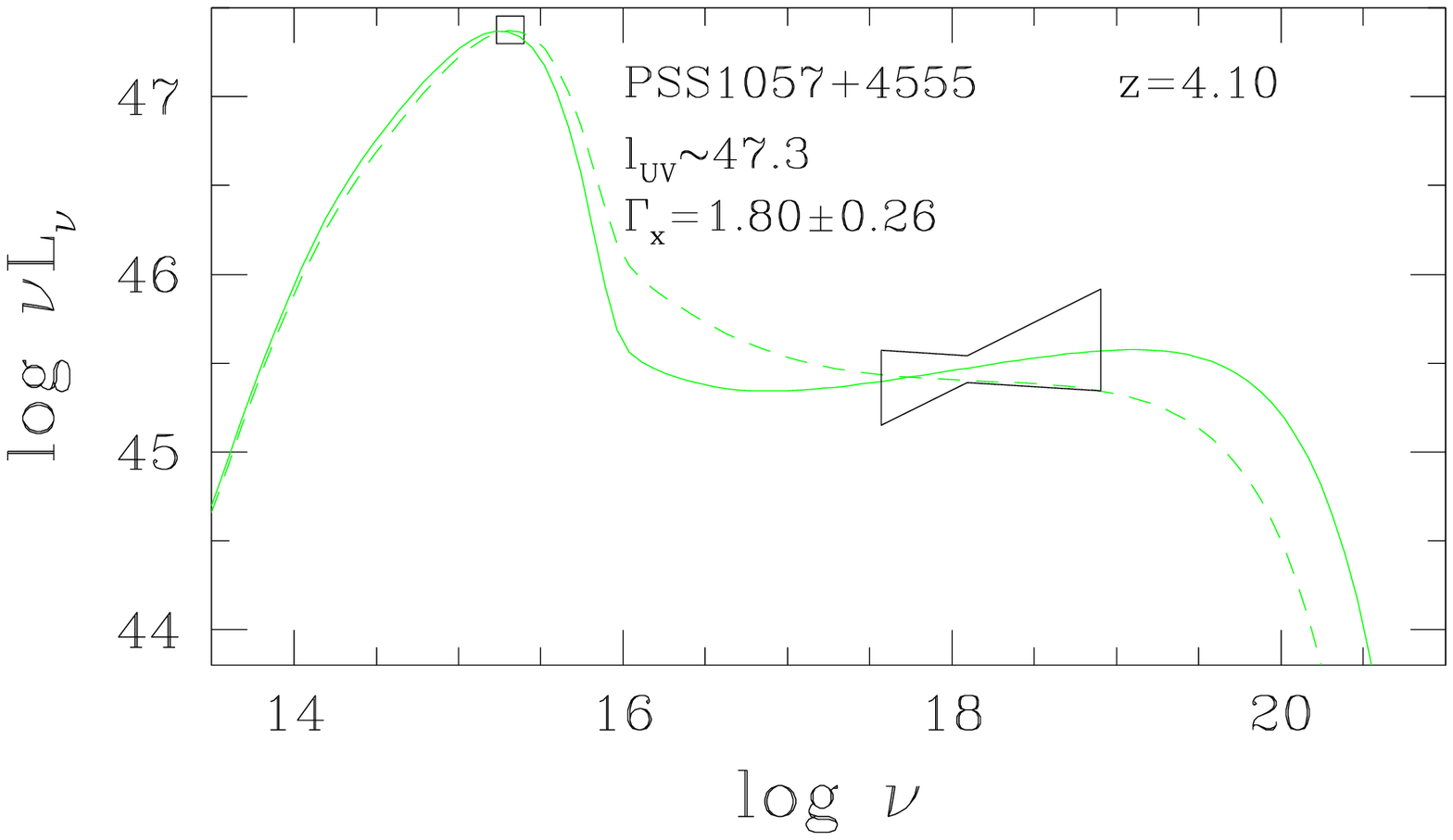}
\caption{Modeling the objects from the B03 sample. {\it Left}: Hot
semispherical flow ($\delta$=1). {\it Right}: Low-efficiency semispherical
flow ($\delta < 1$). The curves show the following cases: $r_{\rm tr} =
r_{\rm s}$ ({\it solid, long-dashed, dot-dashed curves}), $r_{\rm tr}
> r_{\rm s}$ ({\it dashed and dotted curves}), $f > 0.1$ ({\it
  dot-dashed curves}), and plane-parallel geometry ({\it
  dash-long-dashed curves}). Fit
parameters are listed in Tables \ref{tab:fits1} and \ref{tab:fits2}.} 
\label{fig:exmpls}
\end{figure*}

\begin{figure*}
\epsscale{0.45}
\plotone{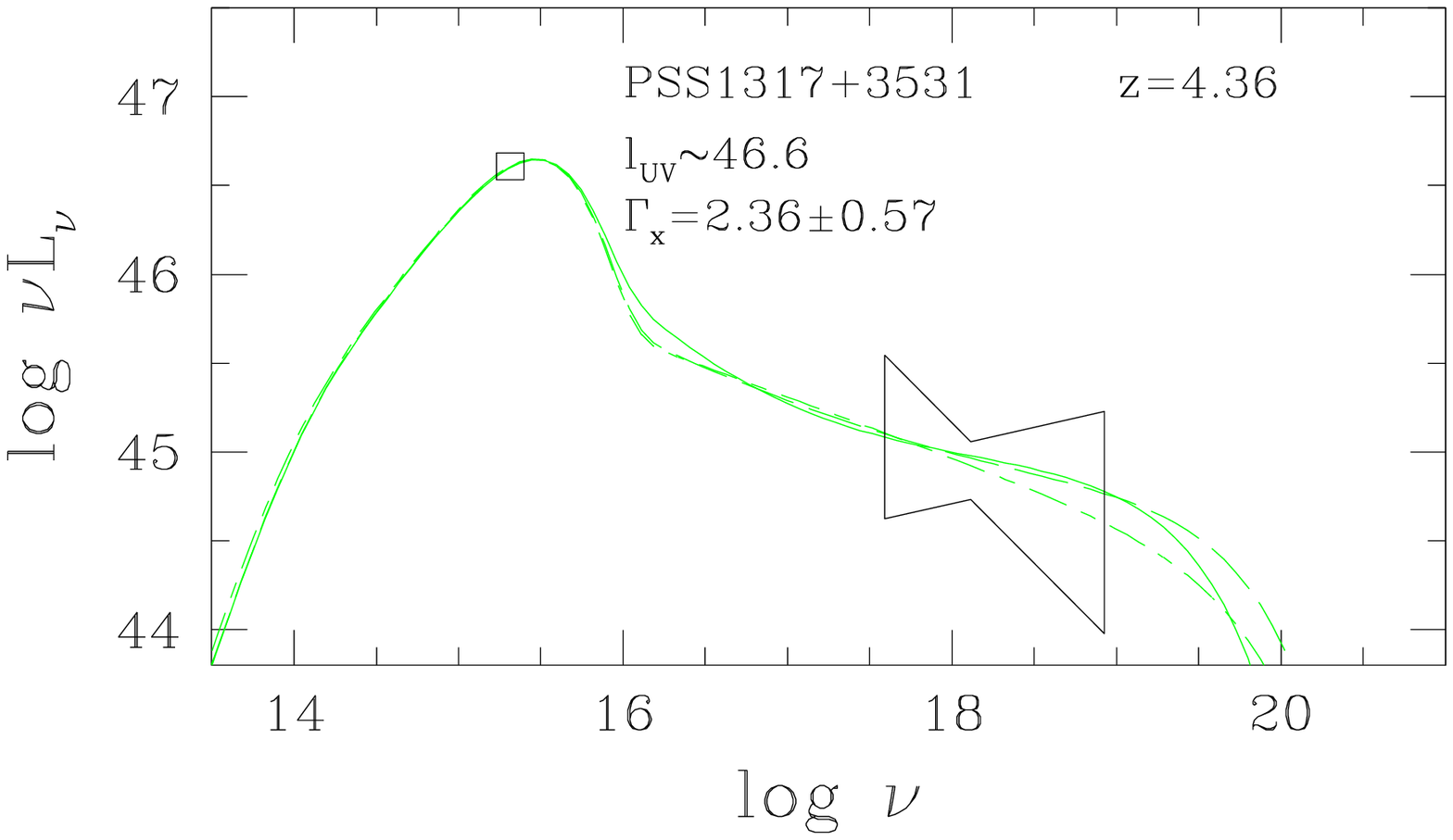}
\epsscale{0.45}
\plotone{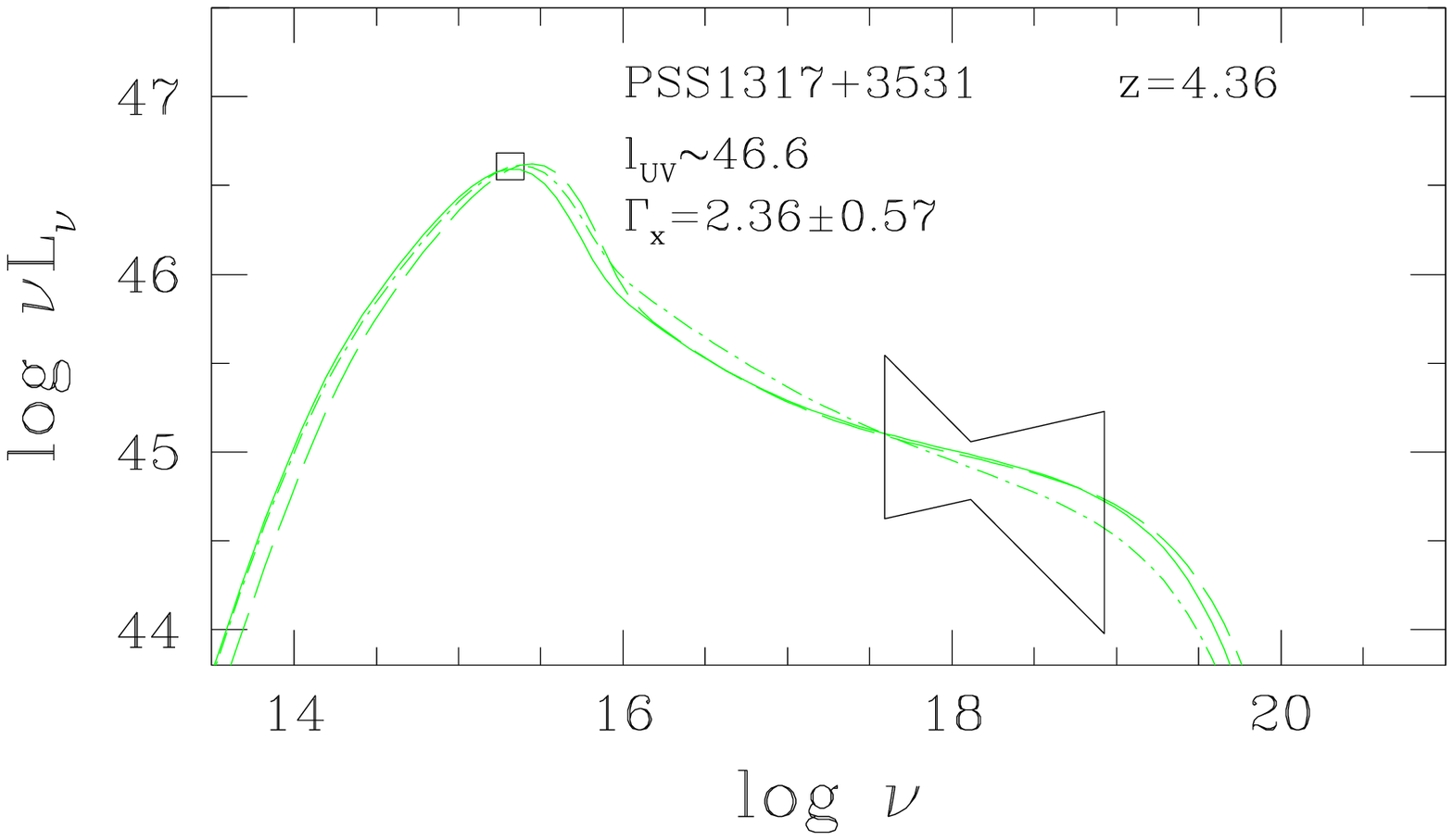}
\epsscale{0.45}
\plotone{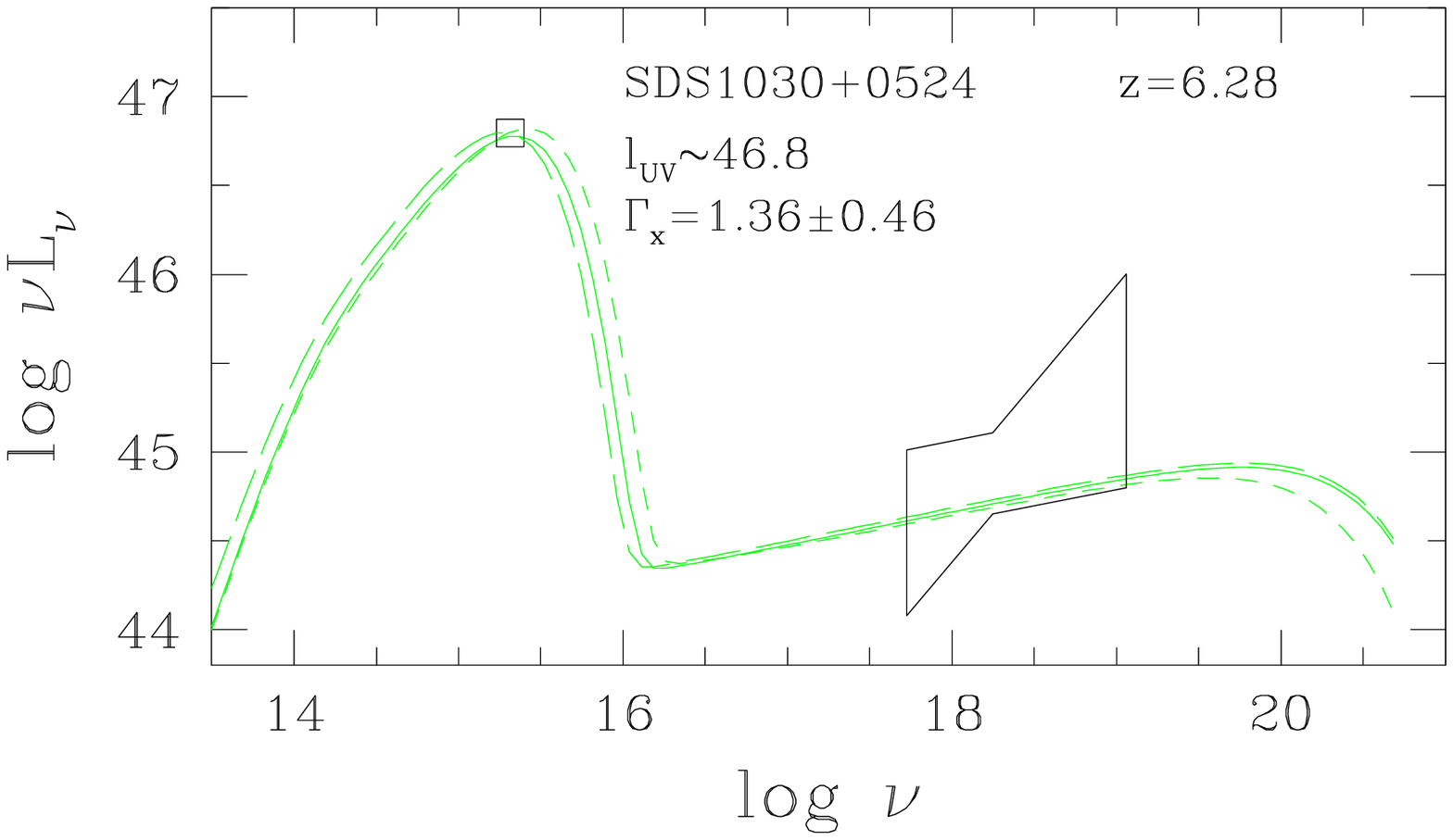}
\epsscale{0.45}
\plotone{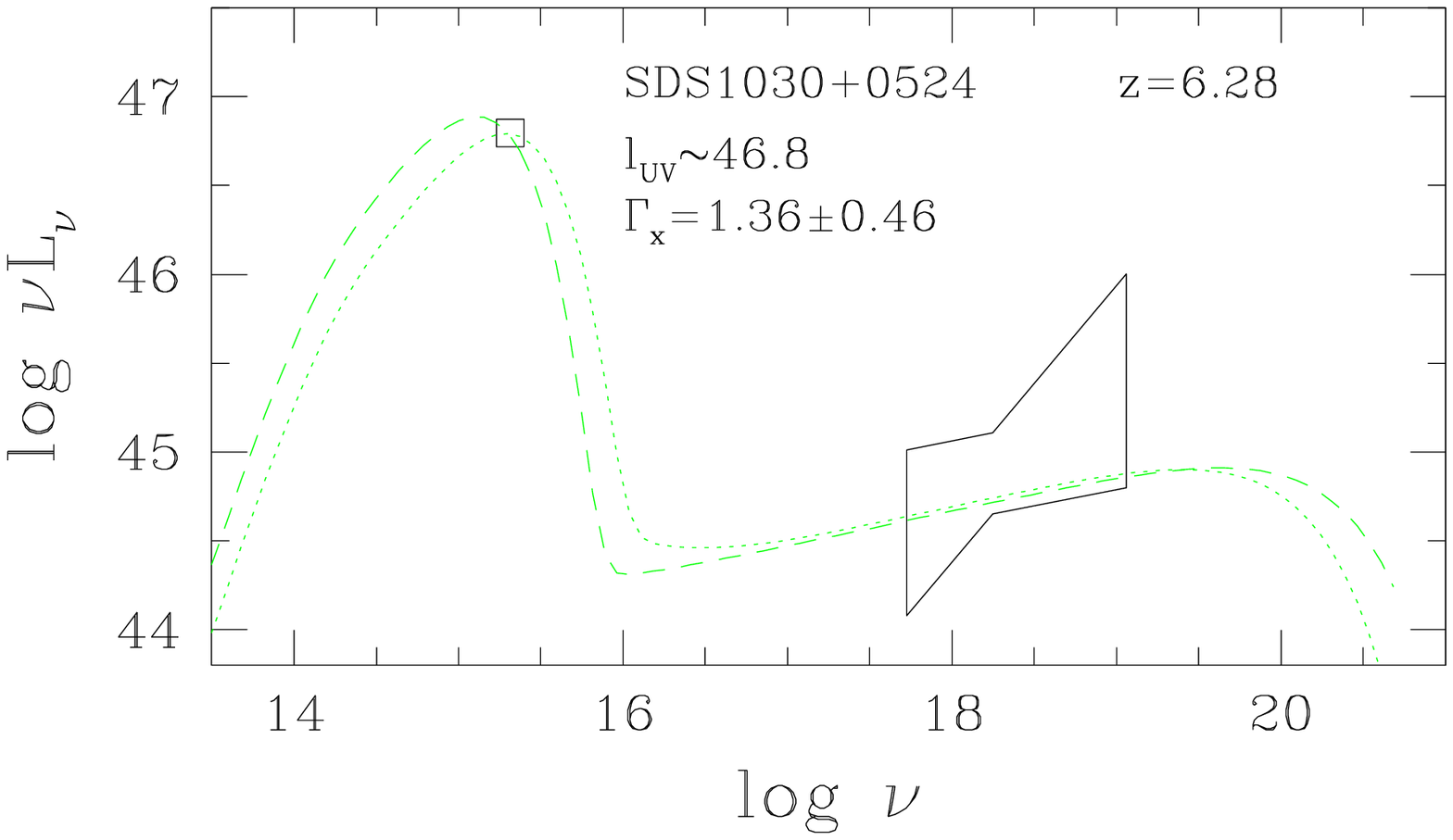}
\epsscale{0.45}
\plotone{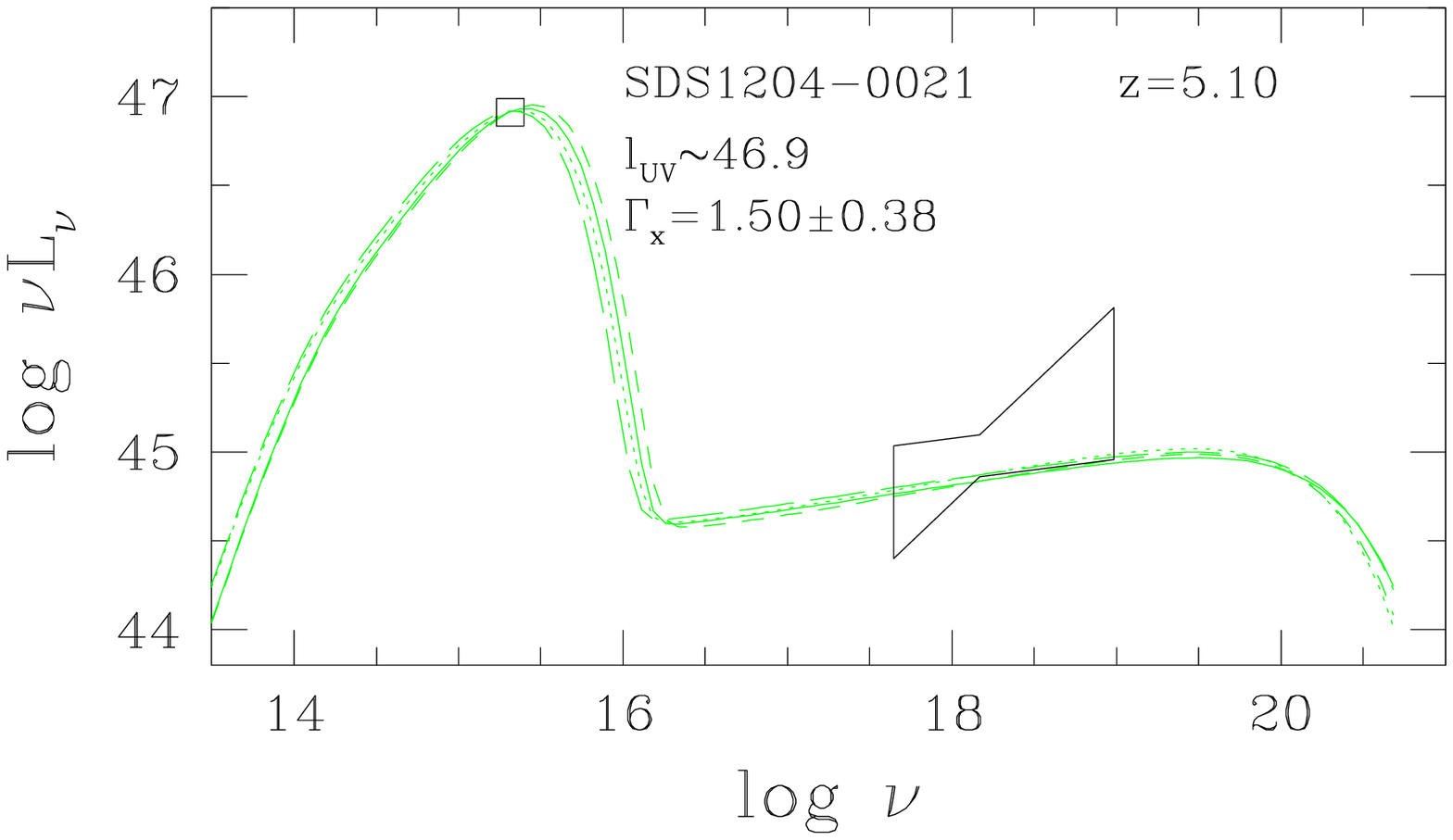}
\epsscale{0.45}
\plotone{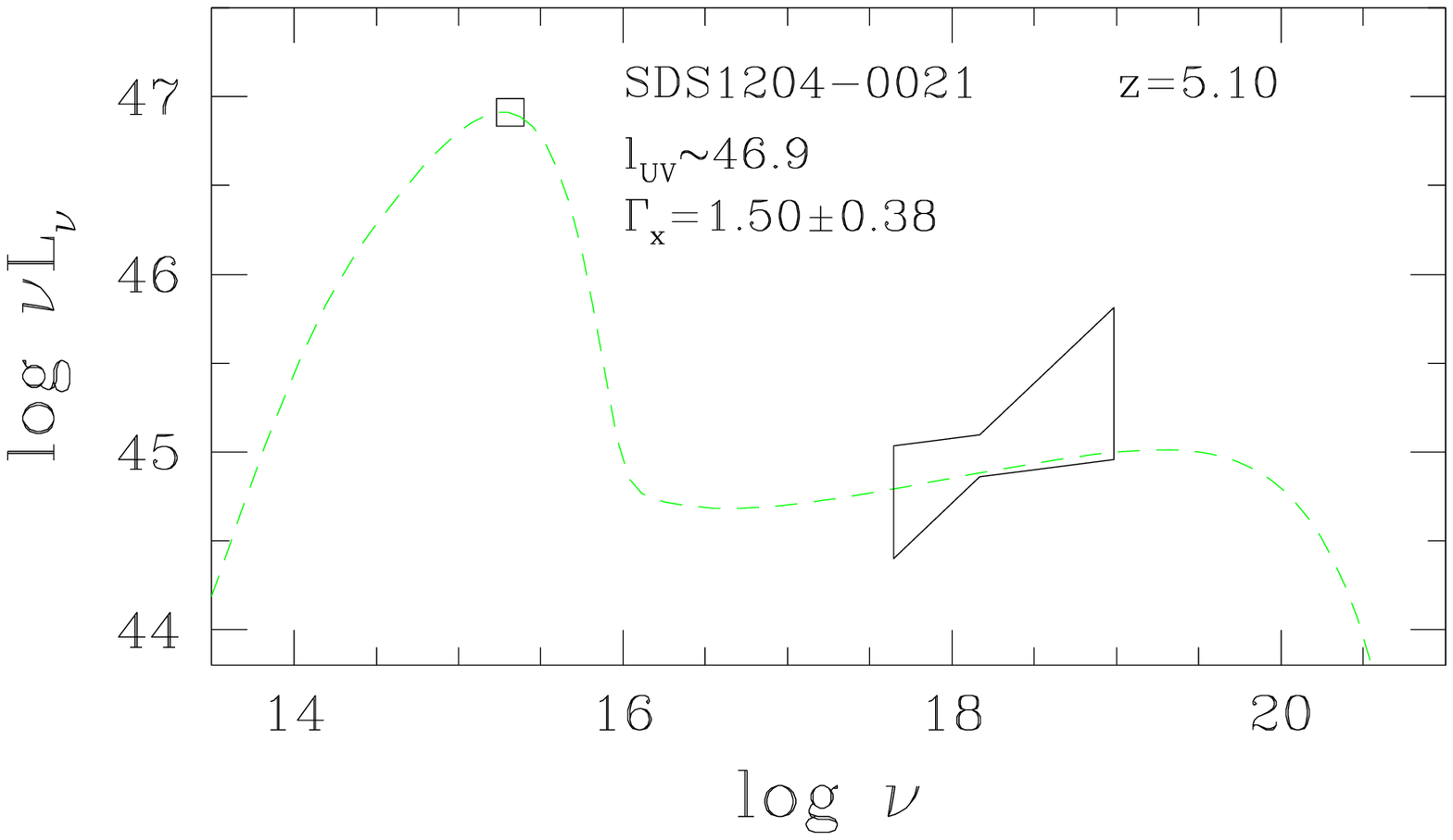}
\epsscale{0.45}
\plotone{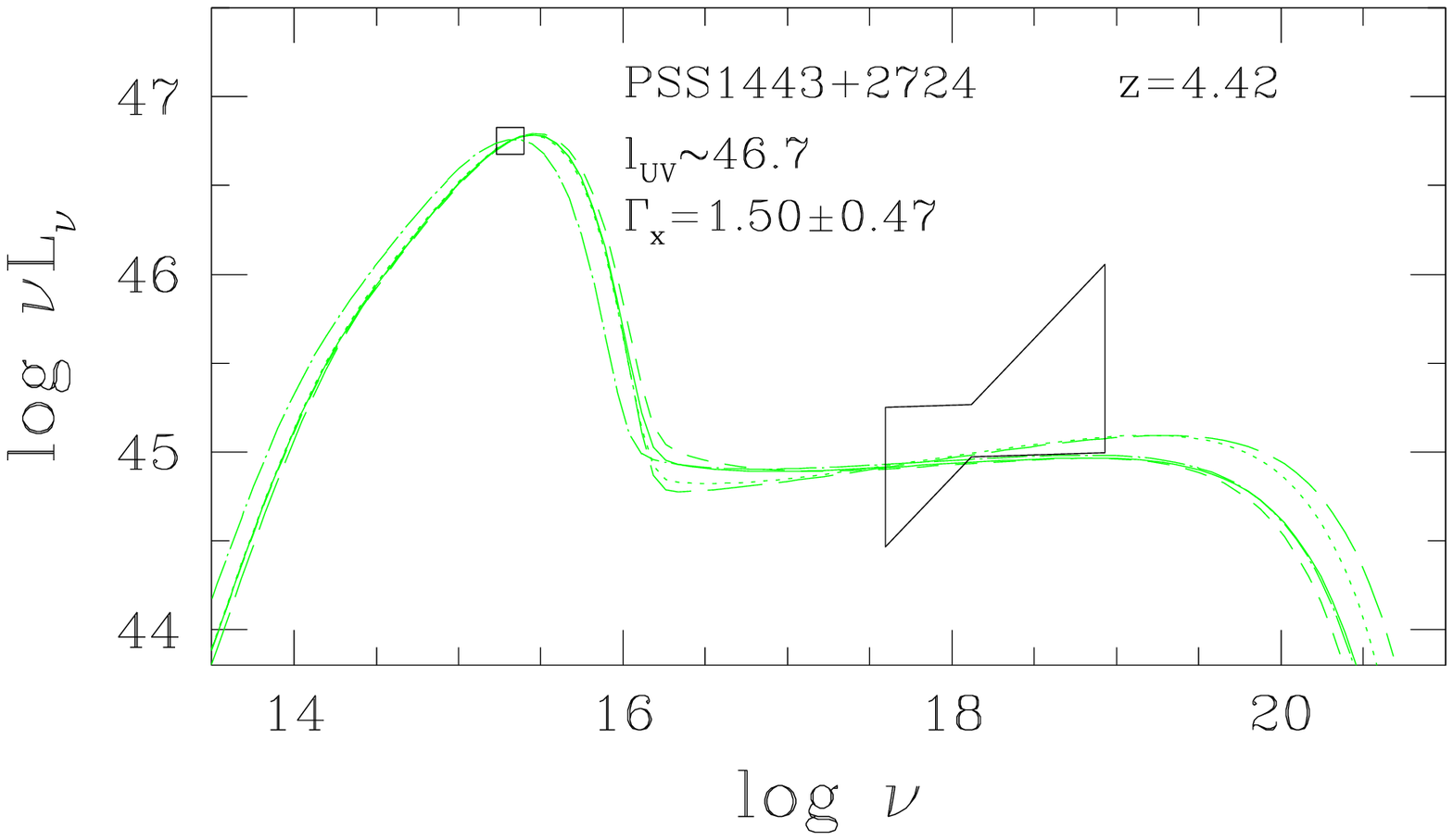}
\epsscale{0.45}
\plotone{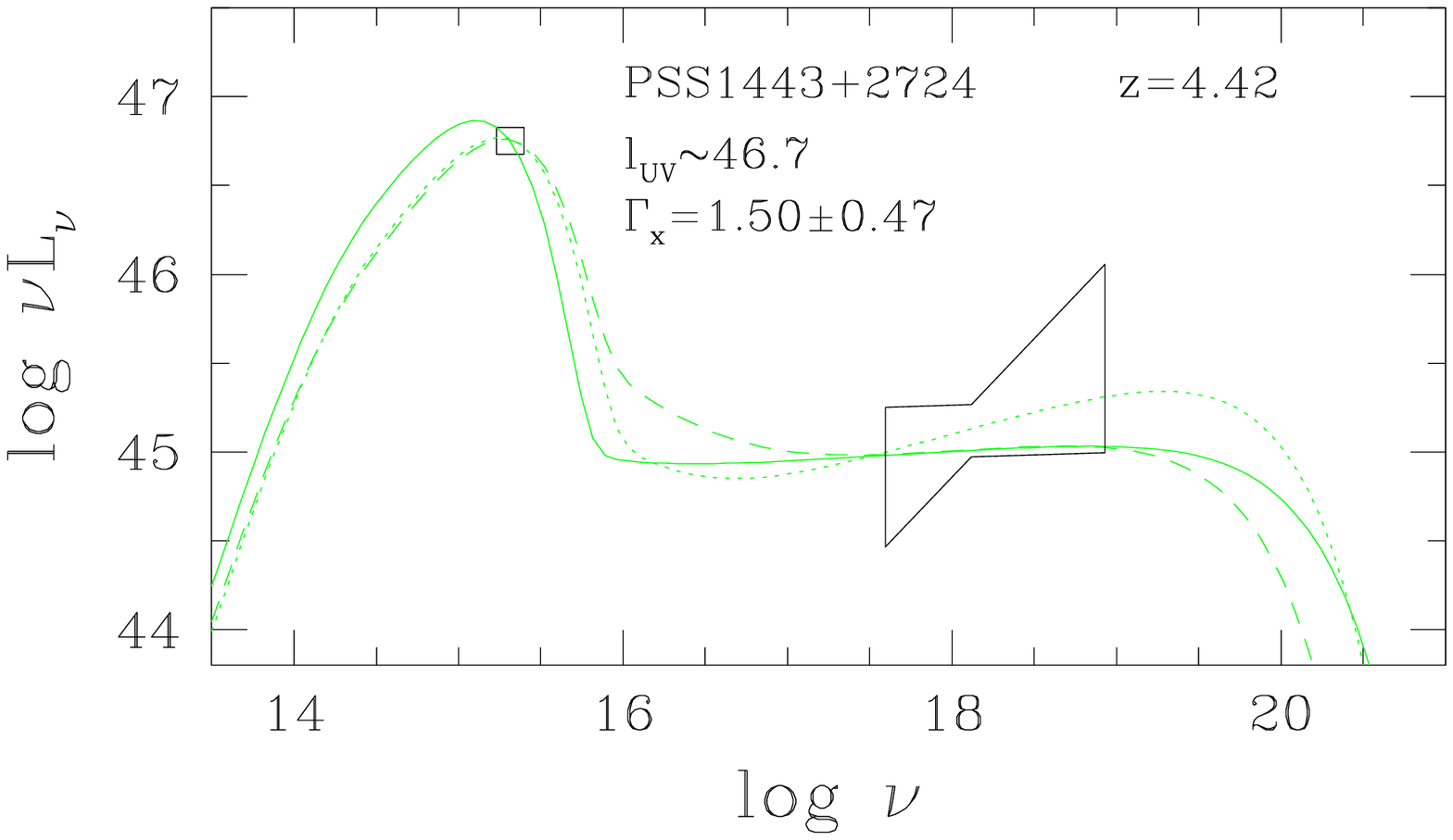}
\caption{Modeling the objects from the B03 sample. {\it Left}: Hot
semispherical flow ($\delta$=1). {\it Right}: Low-efficiency semispherical
flow ($\delta < 1$). The curves show the following cases: $r_{\rm
tr}$=$r_{\rm s}$ ({\it solid, long-dashed, and dot-long-dashed curves}), $r_{\rm tr} > r_{\rm s}$
({\it dashed and dotted curves}), $f > 0.1$ ({\it dot-dashed curves}),
and plane-parallel geometry ({\it dash-long-dashed curves}). Fit parameters are listed in
Tables \ref{tab:fits1} and \ref{tab:fits2}.}
\label{fig:exmpls_cont}
\end{figure*}

We apply the model to eight high redshift RQQ from the BO3
sample. We choose only the seven sources with observed photon index
$\Gamma > 1.5$. We also include the highest redshift quasar, at
z=6.28. We do not perform a formal $\chi^2$ fitting to determine 
the best-fit parameters and their confidence limits, but rather
demonstrate the parameter values that approximately reproduce the spectra.
With the available data the results are not unique, and a number of solutions
are possible. The computed spectra are shown in Figures \ref{fig:exmpls} and
\ref{fig:exmpls_cont}. The fit parameters are listed in Tables 
\ref{tab:fits1} (efficient inner flow) and \ref{tab:fits2} (low
efficiency inner flow). The objects were observed by {\it Chandra} in
the 0.3--6.5 keV range. The 1450 \AA\ rest 
frame points were taken from the literature, as described in detail in
B03. 
The ratio of the optical/UV and X-ray fluxes is characterized by the
X-ray loudness. In the case of high-$z$ quasars, the 2500 \AA\ flux
used to calculate $\alpha_{\rm ox}$  must be extrapolated from the 
1450 \AA\ flux known from observations, which involves a knowledge of the
optical/UV spectral index ($f_{\rm \nu} \sim \nu^{\rm \alpha_{\rm
    UV}}$). To derive general properties of a 
theoretical model or an observational sample, a mean value suggested
by observations can be used (as in Section \ref{sec:results} or in B03,
where $\alpha_{\rm UV} = -0.3$ was adopted; Kuhn et al.\ 2001)). However, it
does not apply to modeling the spectra of particular objects. Thus, in this
section we use the flux at 1450 \AA\ to normalize the spectrum, and
we list values of the spectral index between rest frame 1450 and 2500 \AA,
$\alpha_{\rm UV}$, and the X-ray loudness, $\alpha_{\rm ox}$, computed
from the model spectra in Tables \ref{tab:fits1} and
\ref{tab:fits2}. The optical and X-ray observations were not
simultaneous. However, Giveon et al.\ (1999) 
argue that the optical variability anti-correlates with the luminosity
of AGNs. Given that the high-$z$ RQQs are very luminous objects, one
would expect optical variability at a relatively low level.

Matching the observed optical/UV luminosity requires weak coronae
above the accretion disk, dissipating no more than $\sim 10$\% of the
gravitational potential energy. In addition, high masses of 
the central black holes should be considered, $M \sim (0.5-1.4) \times
10^{10}\, M_{\odot}$. 
Our modeling of BRI0103+0032 illustrates the following trend:
an increase of the black hole mass (with the X-ray slope kept
constant) forces the accretion rate to decrease in order to fit the
optical/UV luminosity. In addition, the electron temperature drops (the
plasma optical depth increases) to match the X-ray emission. 


For the radiatively
efficient spherical inner flow (Section \ref{sec:eff}; $\delta = 1$ and
$r_{\rm s} = r_{\rm tr}$), the truncation radius of the accretion disk
is typically 5--11$\,R_{\rm S}$. This is constrained mainly by the
observed X-ray slope. In general, the electron temperatures
of $\sim 108$--500 keV are required in order to match the observed
X-ray loudness.
The X-ray quietest objects in our sample ($\alpha_{\rm ox}
\sim 1.7-1.8$; PSS1057+4555, SDS1030+0524, and SDS1204-0021) require  very low 
optical depth of the plasma ($\tau < 0.05$), which
brings the temperature to unrealistically high values of 700--1400 keV.
Thus, models with $\delta=1$ are not acceptable, at least for
PSS1057+4555, SDS1030+0524, and SDS1204-0021.

Considering a flattened hot inner flow allows for slightly lower
electron temperatures. However, in the case of objects with
$\alpha_{\rm ox} \sim 1.7$--1.8, the electron temperature is still
high, ranging from 410 to 800 keV. In the case of the two objects with
$\Gamma > 2.3$, there is no need to introduce this modification, since
the flattening of the hot flow would cause the hardening of the
spectrum. These soft objects can be modeled with a very small hot
inner flow (with $r_{\rm tr} \sim 5\, R_{\rm S}$) or even with a
standard plane-parallel configuration in which the accretion disk
covered with a hot corona extends down to the last stable orbit,
$r_{\rm in}$ (in Table \ref{tab:fits1}, fits with $r_{\rm tr} = r_{\rm
s} = r_{\rm in}$). In addition, the corona may be stronger and can dissipate
up to 70\% of the energy (see fits for BRI1033-0327 and PSS1317+3531 in
Table \ref{tab:fits1}).

Assuming the low efficiency spherical flow (Section \ref{sec:loweff};
$\delta < 1$, $r_{\rm tr}=r_{\rm s}$) results in an increase of the disk 
truncation radius needed to fit the observed X-ray slope. Now $r_{\rm
tr} \sim 9 - 35\, R_{\rm S}$ (see Table \ref{tab:fits2}). The electron
temperature needed to match the observed 
X-ray loudness is $kT_{\rm e} \sim 90$--345 keV. However, the objects
with the hardest X-ray spectra (SDS1030+0524 and SDS1204-0021) cannot
be modelled using low efficiency spherical 
flow geometry. Reproducing their X-ray slope (within a 1$\sigma$
confidence interval) requires the disk to be truncated at $60 - 80\,
R_{\rm S}$. With this large an $r_{\rm tr}$, the black hole mass would
have to be larger than $2 \times 10^{10}\, M_{\odot}$ in order to fit
the optical/UV luminosity, even assuming accretion at the
Eddington rate. Even then the fit is not acceptable, because of the
optical/UV spectral index (we discuss this issue in more detail below). 

We also perform fits in which we allow for both low efficiency and
flattening of the hot inner flow. In this approach, all the objects
with $\Gamma  < 2.3$ can be modeled. The disk truncation radius is
$r_{\rm tr} \sim 11$--20$\, R_{\rm S}$, the semi-minor axis of the inner
flow is $r_{\rm s} \sim 4$--10$\, R_{\rm S}$, and the electron
temperature ranges from $\sim 195$ to $\sim 500$ keV. 

The model covers a wide range of the optical/UV spectral index. For
the efficient inner flow, we get $-0.6 < \alpha_{\rm UV} < -0.3$
and for low efficiency inner flow, $-1.5 < \alpha_{\rm UV} <
-0.3$. These values are in agreement with observations that 
suggest a significant spread in $\alpha_{\rm UV}$; e.g., Fan et al.\
(2001) derive $\alpha_{\rm UV} = -0.79 \pm 0.34$ ($1\sigma$), and
Pentericci et al.\ (2003) report on $\alpha_{\rm UV} =  -0.57 \pm
0.33$ ($1\sigma$) for high-$z$ quasars ($3.6 < z < 5$). Knowing the
exact value of $\alpha_{\rm UV}$ for individual sources would reduce
the number of model parameters and make the fitting procedure unique.

We stress that the results for the plasma temperature and optical depth come
from the constraint placed by the ratio of optical/UV to X-ray
emission. The values of X-ray loudness we obtain from fits differ by up
to 16\% from those given by B03. This difference comes from the
adoption of $\alpha_{\rm UV} = -0.3$ and $\Gamma = 2.2$ by B03 when calculating 
$\alpha_{\rm ox}$. 

Our modeling implies a rather high accretion rate, $\dot{M} \sim
0.2$--1.4$\, \dot{M}_{\rm Edd}$. This would suggest that the high-$z$
RQQs may correspond to high or very high states of black hole binaries
(see the comparison of AGN and XRB luminosity states in Czerny
2003). However, the accretion disk in the high or very high state of an XRB
is expected to extend to the last stable orbit (e.g., Esin,
McClintock, \& Narayan 1997).\\

\section{Discussion}
\label{sec:disc}

We have computed optical/UV/X-ray spectra from a two-component accretion flow
in the geometry with a hot, semispherical inner flow enclosed by a
truncated accretion disk covered with a hot corona. We assumed  that the
soft disk blackbody-like radiation 
was Comptonized in the hot plasma and that the electron energy distribution in
the hot plasma was thermal. We investigated the values of model
parameters for which the model can account for spectra observed
in high-redshift RQQs. Such spectra can be characterized
by three observables: (1) the luminosity at 1450 \AA\ in the rest frame of
the source, $l_{\rm UV} \equiv \log \left ( \nu L_{\nu} \right )
\gtrsim 46.0$--47.0, (2) the X-ray loudness, $\alpha_{\rm ox} \sim
1.5$--1.8, and (3) the X-ray photon index, $\Gamma \sim
1.7$--2.3. Finally, we applied the model to the RQQ data of
B03. 

The high luminosities of objects in our sample require not only rather large 
black hole masses, $\sim 10^{10}\,\MSun$, but also high accretion rates,
above $0.2\, \dot{M}_{\rm Edd}$.

We found that some objects require the radiative efficiency of the
hot inner flow to be reduced compared to that of the standard Keplerian disk. A significant
fraction of the energy is then advected inwards, rather than
radiated (i.e., an ADAF-like flow). The hardest ($\Gamma \leq 2.0$) and X-ray
quietest ($\alpha_{\rm ox} \sim 1.7$--1.8) spectra require a
truncation radius up to $r_{\rm tr}\sim 30$--40\, $R_{\rm S}$ if the 
inner flow is spherical. If the inner flow is additionally flattened,
the truncation radius is smaller, yielding $\sim 15$--30\, $R_{\rm S}$.

These values of the truncation radius are comparable to those obtained by
Janiuk et al.\ (2004), who studied a time evolution of unstable accretion disks
with possible disk evaporation. Janiuk et al.\ considered the effects of
the MHD turbulence on the viscosity during the evolution of the
thermal-viscous ionization instability in the standard $\alpha$-accretion
disks, in order to explain intermittent activity in AGNs. They considered the
possibility of accretion through a truncated disk whose inner part is
evaporated to an ADAF-like flow. They found that the evaporation is
also important during the outbursts, even at relatively high accretion
rates of $0.1\, \dot{M}_{\rm Edd}$. The disk then truncates at 
$\sim 25\, R_{\rm S}$
(for a black hole mass of $10^8\, M_{\odot}$ and viscosity parameter
$\alpha = 0.1$). We note, however, that their  value for the disk truncation
radius depends strongly on the adopted value of the viscosity
parameter. 

We were not able to model objects with very hard X-ray spectra
($\Gamma = 0.33$--1.44 in the B03 sample). Such hard
spectra can be produces within the model for the accretion disk
truncated far away from the black hole ($r_{\rm tr} >> 10^2\, R_{\rm
  S}$). However, the optical/UV part of the spectrum is then suppressed
and the model $\alpha_{\rm ox}$ value is much smaller than that observed.

The truncation radius can, in principle, be tested using the X-ray reflection 
component. In the considered geometry, most of the X-ray reflection
comes from disk radii close to $r_{\rm tr}$, so the sharp spectral
features (Fe K$\alpha$ line and edge) should be broadened by the
relativistic effects (Fabian et al.\ 1989; Laor 1991). The amount of
reflection is given by the amplitude of the reflected component,
$R$. In our approach, the accretion disk may reflect hard X-rays
emitted by both the hot corona and the hot inner flow. The amplitude
of reflection is thus given in our model by 
\be
R = \frac{L_{\rm inc}}{L_{\rm hard}} \nonumber,
\ee
where $L_{\rm inc}$ is the luminosity emitted by the hot plasma (hot inner
flow + hot corona) toward the accretion disk and $L_{\rm hard}$ is the
total luminosity of the hot plasma (hot inner flow + hot corona) as seen by a
distant observer, i.e.,
\be
L_{\rm inc} = L_{\rm sphere, inc} + L_{\rm corona, inc} = 2 \pi
\int_{r_{\rm tr}}^{\infty} r F_{\rm inc}(r) dr + 2 \pi \int_{r_{\rm
tr}}^{\infty} r F_{\rm visc}(r) \eta f dr \nonumber
\ee
(we take $2\pi$, since we
consider a top view scenario in which the observer sees just one surface of 
the disk), and 
\be
L_{\rm hard} = L_{\rm diss} + 2 \pi
\int_{r_{\rm tr}}^{\infty} r F_{\rm visc}(r) (1 - \eta) f dr. \nonumber
\ee 
The
reflection amplitude of only the coronal radiation yields $R = 1$, if
one assumes isotropic Comptonization with $\eta=0.5$ (Haardt \&
Maraschi 1991). However, in our geometry the dominant contribution
comes from the reflection of the inner flow radiation by the disk. Thus, the
typical values of the amplitude of reflection are $R = 0.2$--0.6 (the
efficient, spherical inner flow) and $R = 0.15$--0.35 (low-efficiency,
semispherical inner flow). The lower value of the reflection
amplitude in the case of the low-efficiency, semispherical inner flow is
caused by the higher value of the disk truncation radius needed to fit the
data (see Table \ref{tab:fits2}). The higher truncation radius (i.e.,
the accretion disk is truncated farther
away from the black hole) means a smaller contribution from the
reflection of coronal radiation. We note that if the $R - \Gamma$
correlation observed in XRBs and Seyfert galaxies (Zdziarski,
Lubi\'nski, Smith 1999) also holds for high-$z$ RQQs, this model may
account for it. In the case of objects with high accretion rates, the
reflecting material will be highly ionized. As a result, the 
reflected component at energies smaller than the fluorescent iron line energy 
will be flatter, as compared to neutral reflection. Given the small
amplitude of the reflected component calculated above, we estimate that
the change of the X-ray slope due to reflection will be not bigger
than 0.1--0.2, which is well within observational errors. Tests
involving studies of the reflected continuum and iron fluorescence
line require X-ray data of good quality, with the uncertainties not
exceeding 1\%, which may be available in the future observations
with, for example,{\it ASTRO-E2} and the {\it Constellation-X} mission.

\begin{figure*}
\epsscale{0.45}
\plotone{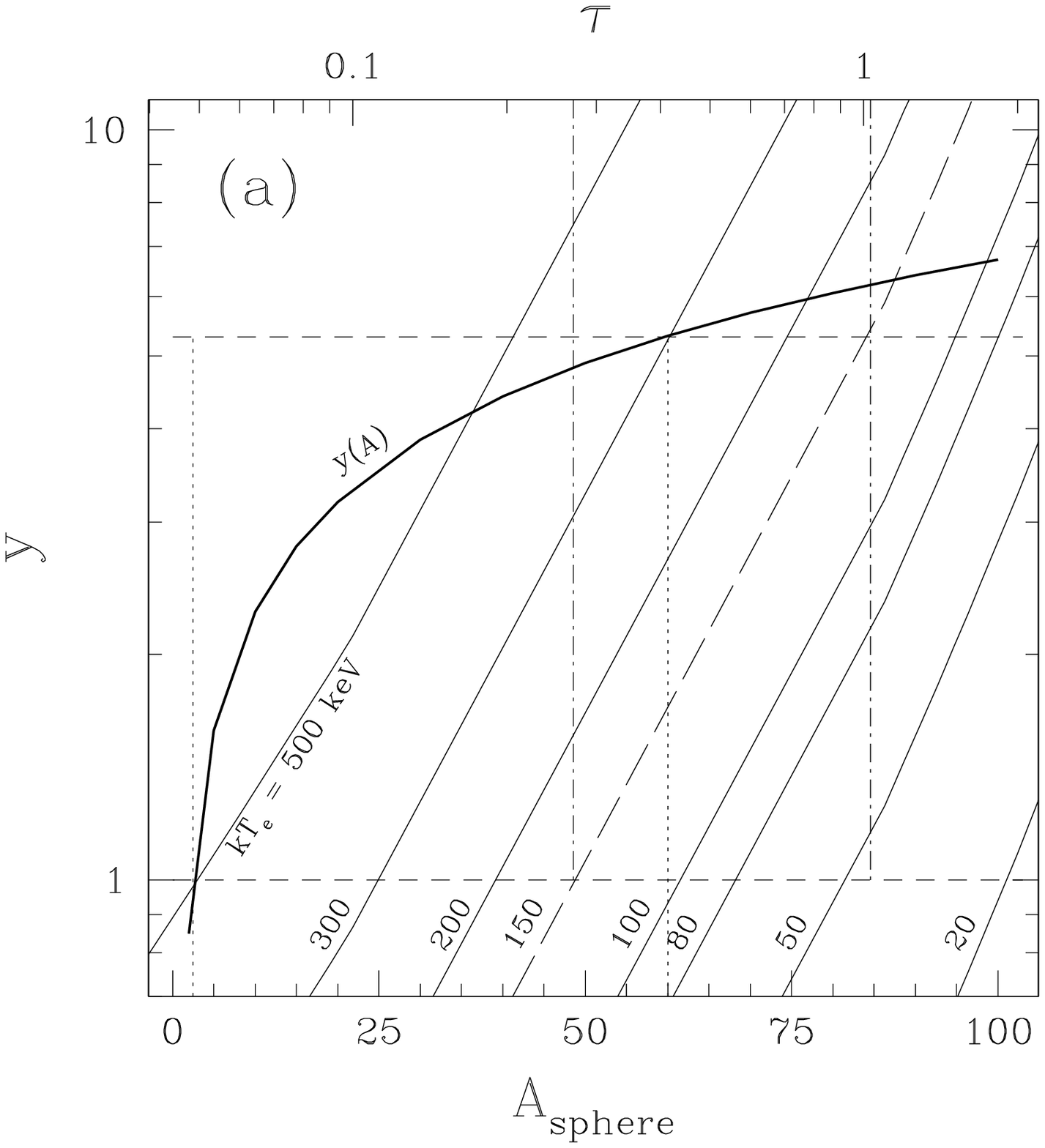}
\epsscale{0.45}
\plotone{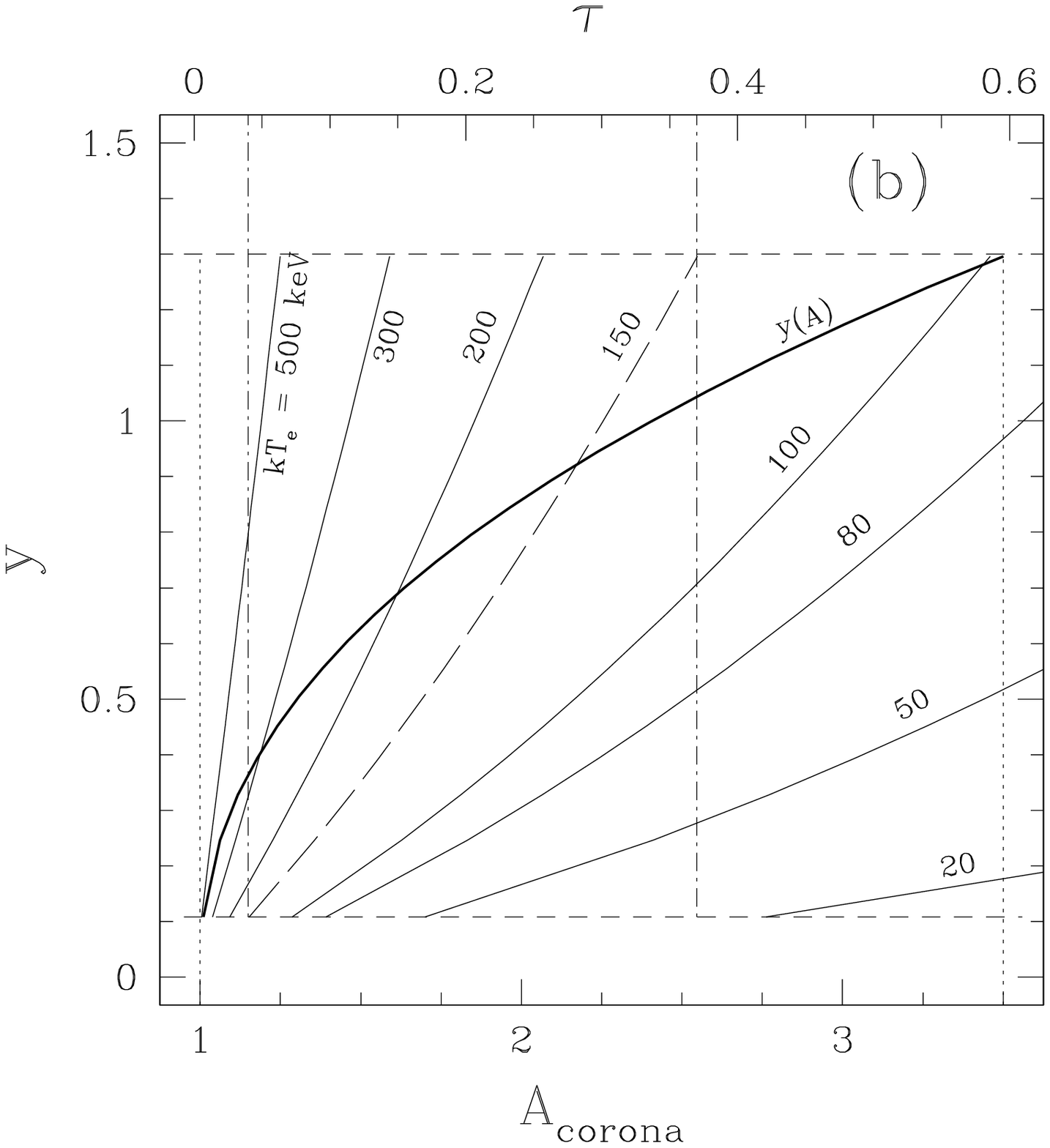}
\caption{Hot plasma parameters in the model with no reduction in efficiency
($\delta = 1$), and spherical inner  flow ($r_{\rm tr}=r_{\rm s}$) for (a)
$r<r_{\rm tr}$ (the inner flow) and (b) $r>r_{\rm tr}$ (the 
corona). The thick solid curves show the dependence of the Compton
parameter, $y$, on the amplification factor, $A$, obtained by combining 
eqs.~(\ref{b1}) and (\ref{b2}). Vertical dotted lines (at [a] $A=2.5$ and
$A=60$ and [b] $A=1$ and $A=3.5$)
show the range of $A$ possible to obtain from
the model. The horizontal dashed lines show the corresponding range of $y$. 
Solid lines illustrate the relation between the
Compton parameter, $y$, the optical depth, $\tau$ ({\it upper axis}), and the electron 
temperature, $kT_{\rm e}$ (eq.~[\ref{equ:ypar}]). The range of $y$ marked by
the horizontal lines, corresponds to the range of $\tau$ for a given 
$kT_{\rm e}$. For illustration, the range of $\tau$ for $k T_{\rm e}=150$ keV 
is shown by dash-dotted lines.}
\label{fig:yatau}
\end{figure*}

Results of our modeling imply that the temperature of Comptonizing
plasma should be rather high, up to 500 keV. This
constraint comes from the required low optical depth of the
Comptonizing plasma needed to reproduce the relatively low
normalization of hard X-rays compared to that of the disk component 
(i.e., the high value of $\alpha_{\rm ox}$).  In Figure
\ref{fig:yatau}, we present the range of the amplification factor that
we obtain for the Comptonization in the efficient, hot, spherical inner
flow ($A = 2.5$--60, lower axis) and the corresponding range of the
optical depths ($\tau = 0.3$--1.0; upper axis) for the electron
temperature 150 keV.  Similar dependencies are shown for the
Comptonization in the corona ($A = 1.0$--3.5; $\tau =
0.05$--0.37). Obviously, the higher the electron temperature, the
lower the optical depth, because the X-ray spectral slope is fixed by the
data.

Fits to the spectra of accreting black hole XRBs in the hard state
and local Seyfert galaxies give electron temperatures of 50--150 keV and
optical depths of the order of $\tau \sim 1$
(Gierli\'nski et al.\ 1997; Zdziarski et al.\ 1998; Zdziarski, Poutanen 
\& Johnson 2000; Done et al.\ 2000). In particular, Zdziarski et al.\ (2000)
analyzed the average spectrum of 17 radio quiet Seyfert 1 galaxies in the
50--500 keV range  observed with OSSE. They chose the spherical geometry
with an input of soft photons at the temperature of 10 eV in the center of the
sphere. They used the {\tt compps} model in the {\sc Xspec} package to fit
the Comptonized continuum 
together with the reflection feature. Thus, the high energy cutoff in the model
spectrum, which determines the plasma temperature as a physical parameter,
was directly constrained by the data points. Their derived plasma
temperatures for Seyfert galaxies were much lower than the
temperatures required for high redshift quasars in our modelling.
 
On the other hand, Chiang (2002) and Chiang \& Blaes (2003) performed
analysis of three Seyfert galaxies that was very similar to the
analysis in the present paper. By fitting the spectra including both the
optical/UV and the 2--20 keV continuum, they also obtained rather high
plasma temperatures (sometimes above 500 keV) and low optical
depths. The difference between 
their approach and the approach of Zdziarski, Poutanen \& Johnson (2000)
was modeling the X-ray data together with optical/UV data. In addition,
Chiang (2002) and Chiang \& Blaes (2003) attempted to additionally constrain 
their model by inferring black hole masses and disk truncation radii from other
observations (reverberation mapping, width of the Fe K$\alpha$ line, etc.).
They did not use data above 30 keV, so the hard X-ray 
cutoff was a free parameter in their model, as it was in ours.

To qualitatively compare high- and low-$z$ objects, we perform
exemplary fits to the Seyfert galaxy NGC 7469 (analyzed by Chiang
2002) and the low-$z$ (z=0.206) quasar PG 0947+396 from a sample
described in Laor et al.\ (1997). For NGC 7469 we obtain a good fit
for $kT_{\rm e}=110$ keV and $\tau_{\rm s}=0.8$ (in agreement with
findings of Zdziarski et al.\ 2000), a truncation radius of $r_{\rm
  tr} = 27\, R_{\rm S}$, an accretion rate of 0.17 times the Eddington rate,
and a mass of $5{\times}10^7\, M_{\odot}$. The inner flow efficiency is
low, comparable to that of Chiang (2002). In the case of PG 0947+396, we
require the accretion disk to extend to 
the last stable orbit, the corona to dissipate 40\% of the energy, the
plasma temperature $kT_{\rm e}$ to be 230 keV, and the optical depth $\tau_{\rm
  c}$ to be 0.1. The mass estimate is 
$5{\times}10^8\, M_{\odot}$, and the accretion rate accounts for 0.7
times the Eddington rate. These results suggest that low-$z$ objects
can be modeled with
lower masses and different plasma parameters (lower temperatures and
higher optical depths). However, a satisfactory comparison between low-
and high-$z$ samples cannot be made at 
this stage. First, the X-ray data do not correspond to the same rest
frame energies: 2--30 keV for high-$z$ quasars, 2--10 keV for NGC 7469
(Nandra 2000), and 0.3--2 keV for PG 0947+396 (Laor et al.\
1997). Second, the low-$z$ objects require very detailed data analysis
in the 0.3--10 keV X-ray range to model the soft X-ray excess, and above 10
keV to model hard X-rays. In principle, our model is able to explain
the soft excess in terms of two Comptonizations, since it includes two
Comptonizing media: a corona above the accretion disk and an inner
flow, unlike, e.g., the models of Chiang (2002) and Chiang \& Blaes
(2003). However, in the recent paper of Gierli\'nski \& Done (2004), the
authors model the 0.3--20 keV range with only one Comptonization
component and a complex absorption feature and point out that the
soft X-ray excess may be an artifact of such an absorption. The
present data of high-$z$ quasars do not allow a determination of whether
their rest-frame spectra show the soft X-ray excess or not, because the
component is redshifted below the observable band. 
On the other hand, some absorption features may become important in the
future, higher quality X-ray data.

Modeling broad-band quasar spectra including both accretion disk
emission and the Comptonized hard X-rays provides a difficult
theoretical challenge. Reproducing the relative normalization of the
two main emission components requires rather detailed knowledge of the
geometry and emission mechanisms and patterns. Moreover, the high
accretion rate inferred from our modeling implies that the studied
objects may correspond to high or very high states of galactic black
hole binaries. In addition, the resulting high electron temperatures and low
optical depths in the scenario with a thermal plasma may suggest that
the Comptonization actually takes place in non-thermal or hybrid
plasmas, as is seen in XRBs in their high state (i.e., Gierli\'nski et
al.\ 1999). Many detailed studies of XRBs reveal complex,
multi-component spectra (see, e.g., \.{Z}ycki, Done \& Smith 2001;
Wilson \& Done 2001; Gierli\'{n}ski \& Done 2003). The spectra contain
not only simple disk emission with a hard Comptonized component, but
also additional Comptonization of the disk emission, modification due
to X-ray irradiation of the external regions of the disk, and the
reprocessed component, usually ionized and relativistically
smeared. These additional components are also detected in AGN spectra
(e.g., PG~1211+143; Janiuk, Czerny \& Madejski 2001), and will surely
have to be incorporated into future models (e.g., Czerny et al.\ 2003).\\

\section{Conclusions}
\label{sec:concl}

In this paper we explored a parameter space of the semi-spherical
accretion models applicable to high-$z$ RQQs. Our results show the following:

\noindent
1. The observed optical/UV luminosity suggests that high redshift
quasars host supermassive black holes with masses of the order of
$10^{10} M_{\odot}$. The modeling results in accretion rates higher
than $\sim 0.2$ times the Eddington rate.

\noindent
2. The high accretion rates may suggest similarity between high-$z$ RQQs and
XRBs in the high or very high state. If high-$z$ RQQs 
correspond to high or very high states of X-ray binaries, the high
values of plasma temperature (up to 500 keV) and low optical depths
might suggest instead that Comptonization takes place in a non-thermal
or hybrid plasma. 

\noindent
3. The observed X-ray loudness puts strong constraints on the modeled
plasma temperature and optical depth. {\it Chandra} or {\it XMM-Newton}
observations with better S/N are needed to give tighter constraints on the X-ray slope.
Future mission (such as {\it Constellation-X}) can provide information about the
high-energy cut-off and the reflected component that is necessary to
uniquely determine geometry of the accretion flow. 

\noindent
4. Precise observations constraining the optical/UV
spectral index for individual sources would help to reduce the number
of model parameters and perform more accurate fits. However, the
intrinsic reddening of the quasar spectrum is the main observational
uncertainty in this spectral region. 

\noindent
5. With the present data, a number of theoretical descriptions are possible.
In Paper II, we focus on another interesting possibility, in which the
accretion disk extends to the last stable orbit and the hot Comptonizing
corona is formed by clouds above the disk (the ``patchy'' corona geometry).

\acknowledgements
We thank Martin Elvis, Paul Green, Jill Bechtold, and Jim Chiang for
many discussions. We also thank the anonymous referee for detailed
reading of the manuscript and helpful comments. This research is
funded in part by NASA contract NAS8-39073.  Partial support for this
work was provided by the National Aeronautics and Space Administration
through {\it Chandra} Awards Number GO1-2117B and GO2-3148A, issued by the
{\it Chandra} X-Ray Center, which is operated by the Smithsonian
Astrophysical Observatory for and on behalf of NASA, under contract
NAS8-39073. M.A.S. and P.T.{\.Z}. were partially supported by the
grant PBZ-KBN-054/P03/2001 and KBN project number 2P03D00322. M.A.S.
acknowledges support from the Smithsonian Institution Pre-doctoral
Fellowship program. 

\appendix

\section{Appendix}

For completeness, we give here expressions for 
$g(r)$, $\alpha_{\rm max}$, $\phi_{\rm max}$, and $l(r,\alpha,\phi,r_{\rm
s})$ (see Chiang 2002):
$$g(r)=2\int_0^{\alpha_{\rm max}} \phi_{\rm
max}(\alpha,r)\sin\alpha\cos{\alpha}d\alpha,$$
$$\alpha_{\rm max}=\tan^{-1}\left(\frac{r_{\rm s}}{(r^2-r_{\rm
tr}^2)^{1/2}}\right),$$
$$\phi_{\rm max} = \cos^{-1}\left[ \frac{\left[(r^2-r_{\rm tr}^2)(r_{\rm s}^2 
+ \tan^2 \alpha)\right]^{1/2}}{r r_{\rm s}}\right],$$
$$l = \frac{2r_{\rm s}\left[ (r_{\rm s}^2+\tan^2{\alpha}) -
r^2(\tan^2{\alpha}+r_{\rm
s}^2\sin^2{\phi})\right]^{1/2}}{\cos{\alpha}(r_{\rm s}^2 +
\tan^2{\alpha})}.$$


\begin{deluxetable}{l cccccccc ccc}
\tabletypesize{\scriptsize}
\tablecaption{Modeling the SED of high-$z$ RQQs. Efficient Inner Flow. \label{tab:fits1}}
\tablewidth{0pt}
\tablehead{
 & \multicolumn{9}{c}{$\delta=1$}\\
{\sc Object} & $m$\tablenotemark{a} & $\dot{m}$\tablenotemark{b} & 
$f$\tablenotemark{c} & $r_{\rm tr}$\tablenotemark{d} & 
$r_{\rm s}$\tablenotemark{d} & $kT_{\rm e}$\tablenotemark{e} & 
$\tau_{\rm c}$\tablenotemark{f} & $\tau_{\rm s}$\tablenotemark{f} & 
$\Gamma_{\rm fit}$\tablenotemark{g} & $\alpha_{\rm
  UV}$\tablenotemark{h} & $\alpha_{\rm ox}$\tablenotemark{i}
}
\startdata
BRI0103+0032 & $5.5{\times}10^{9}$ & 0.9 & 0.1 & 10 & 10 & 520 & 0.01 &
0.11 & 1.84 & -0.28 & 1.57\\
~& $1{\times}10^{10}$ & 0.45 & 0.1 & 10 & 10 & 480 & 0.02 & 0.13 &
1.84 & -0.62 & 1.59\\
~ & $8{\times}10^9$ & 0.48 & 0.1 & 7 & 7 & 290 & 0.04 & 0.22 & 1.96
& -0.35 & 1.57\\
~&~&~&~&~&~&~&~&~&~&~&~\\
~& $1{\times}10^{10}$ & 0.35 & 0.1 & 8 & 5.5 & 370 & 0.03 & 0.20 &
1.84 & -0.52 & 1.58\\
~ & $8.5{\times}10^9$ & 0.42 & 0.1 & 6 & 4 & 230 & 0.06 & 0.32 & 1.96
& -0.34 & 1.57\\
\tableline
PSS0248+1802 & $1{\times}10^{10}$ & 0.65 & 0.1 & 7 & 7 & 240 & 0.05 &
0.29 & 1.96 & -0.33 & 1.52\\
~&~&~&~&~&~&~&~&~&~&~&~\\
~ & $1{\times}10^{10}$ & 0.64 & 0.1 & 6 & 4 & 190 & 0.08 & 0.40 & 1.97
& -0.28 & 1.52\\
\tableline
BRI1033-0327 & $1{\times}10^{10}$ & 0.31 & 0.1 & 5 & 5 & 108 & 0.17 & 0.59
& 2.19 & -0.33 & 1.53\\
~& $1{\times}10^{10}$ & 0.54 & 0.7 & 5 & 5 & 375 & 0.08 & 0.14 & 2.19
& -0.34 & 1.53\\
~ & $1{\times}10^{10}$ & 0.44 & 0.56 & $r_{\rm in}$ & $r_{\rm in}$ &
252 & 0.12 &\nodata& 2.39 & -0.29 & 1.53\\
~ & $1.5{\times}10^{10}$ & 0.27 & 0.56 & $r_{\rm in}$ & $r_{\rm in}$ &
230 & 0.14 & \nodata& 2.39 & -0.49 & 1.55\\
~&~&~&~&~&~&~&~&~&~&~&~\\
~&\nodata&\nodata&\nodata&\nodata&\nodata&\nodata&\nodata&\nodata&\nodata&\nodata&\nodata\\
\tableline
PSS1057+4555 & $1.3{\times}10^{10}$ & 1 & 0.1 & 9 & 9 & 700 & 0.01 &
0.06 & 1.86 & -0.41 & 1.70\\
~& $1.2{\times}10^{10}$ & 0.98 & 0.1 & 7 & 7 & 495 & 0.02 & 0.10 &
1.94 & -0.28 & 1.70\\
~&~&~&~&~&~&~&~&~&~&~&~\\
~& $1.2{\times}10^{10}$ & 1 & 0.1 & 7 & 4 & 570 & 0.01 & 0.10 &
1.84 & -0.30 & 1.70\\
~& $1.4{\times}10^{10}$ & 0.78 & 0.1 & 6 & 4 & 410 & 0.02 & 0.14 &
1.93 & -0.32 & 1.70\\
\tableline
PSS1317+3531 & $6{\times}10^9$ & 0.26 & 0.1 & 5 & 5 & 115 & 0.16 & 0.55 &
2.19 & -0.27 & 1.52\\
~& $6{\times}10^9$ & 0.34 & 0.4 & 5 & 5 & 230 & 0.11 & 0.25 & 2.25 &
-0.27 & 1.52\\
~ & $7{\times}10^9$ & 0.32 & 0.61 & $r_{\rm in}$ & $r_{\rm in}$ & 275 & 0.11 
&\nodata& 2.36 & -0.30 & 1.52\\
~ & $1{\times}10^{10}$ & 0.21 & 0.61 & $r_{\rm in}$ & $r_{\rm in}$ &
260 & 0.12 &\nodata& 2.36 & -0.46 & 1.55\\
~&~&~&~&~&~&~&~&~&~&~&~\\
~&\nodata&\nodata&\nodata&\nodata&\nodata&\nodata&\nodata&\nodata&\nodata&\nodata\\
\tableline
SDS1030+0524 & $5{\times}10^{9}$ & 0.79 & 0.1 & 11 & 11 & 1390 & 0.002 &
0.02 & 1.82 & -0.34 & 1.79\\
~ & $1{\times}10^{10}$ & 0.38 & 0.1 & 11 & 11 & 1310 & 0.003 & 0.02 &
1.82 & -0.76 & 1.81\\
~&~&~&~&~&~&~&~&~&~&~&~\\
~& $7{\times}10^{9}$ & 0.45 & 0.1 & 7.2 & 4 & 850 & 0.01 & 0.05 & 1.82
& -0.37 & 1.79\\
~ & $1{\times}10^{10}$ & 0.3 & 0.1 & 7.2 & 4 & 800 & 0.01 & 0.05 &
1.82 & -0.56 & 1.80\\
\tableline
SDS1204-0021 & $7{\times}10^9$ & 0.65 & 0.1 & 9 & 9 & 930 & 0.005 &
0.04 & 1.86 & -0.36 & 1.77\\
~& $1{\times}10^{10}$ & 0.44 & 0.1 & 9 & 9 & 880 & 0.01 & 0.04
& 1.86 & -0.56 & 1.77\\
~&~&~&~&~&~&~&~&~&~&~&~\\
~ & $7{\times}10^9$ & 0.6 & 0.1 & 7 & 4 & 750 & 0.01 & 0.06 & 1.84 &
-0.29 & 1.77\\ 
~ & $1{\times}10^{10}$ & 0.39 & 0.1 & 7 & 4 & 690 & 0.01 & 0.07 & 1.84
& -0.47 & 1.77\\
\tableline
PSS1443+2724 & $5{\times}10^9$ & 0.65 & 0.1 & 9 & 9 & 590 & 0.01 &
0.08 & 1.87 & -0.29 & 1.64\\ 
~& $6{\times}10^9$ & 0.46 & 0.1 & 7 & 7 & 400 & 0.02 & 0.14 &
1.95 & -0.30 & 1.64\\
~& $1{\times}10^{10}$ & 0.26 & 0.1 & 7 & 7 & 380 & 0.03 & 0.15 & 1.95
& -0.56 & 1.66\\
~&~&~&~&~&~&~&~&~&~&~&~\\
~& $6{\times}10^9$ & 0.48 & 0.1 & 7.3 & 5 & 450 & 0.02 & 0.13 & 1.87 &
-0.32 & 1.64\\  
~& $6{\times}10^9$ & 0.44 & 0.1 & 6 & 4 & 330 & 0.03 & 0.19 & 1.94 &
-0.26 & 1.64\\
~& $1{\times}10^{10}$ & 0.25 & 0.1 & 6 & 4 & 310 & 0.04 & 0.21 & 1.94
& -0.52 & 1.66\\
\enddata

\tablenotetext{a}{Black hole mass in Solar masses, $M_{\odot}$}
\tablenotetext{b}{Accretion rate in Eddington units, $\dot{M}_{\rm Edd}$}
\tablenotetext{c}{Fraction of gravitational energy dissipated in the corona above 
the disk}
\tablenotetext{d}{Disk truncation radius,$r_{\rm tr}$, and the semi-minor axis of
the ellipsoidal flow, $r_{\rm s}$, in Schwarzschild units, $R_{\rm S}$}
\tablenotetext{e}{Plasma temperature in keV}
\tablenotetext{f}{Optical depth of corona, $\tau_c$, and semispherical flow,
$\tau_s$}
\tablenotetext{g}{X-ray photon index allowed to vary within the 1$\sigma$ confidence 
interval (we do not list $l_{\rm UV}$ from fits since its value was fixed at that observed)}
\tablenotetext{h}{UV index between 1450 and 2500 \AA\ in the rest frame}
\tablenotetext{i}{X-ray loudness}

\end{deluxetable}

\begin{deluxetable}{l cccccccc ccc}
\tabletypesize{\scriptsize}
\tablecaption{Modeling the SED of high-$z$ RQQs: Low-Efficiency Inner Flow \label{tab:fits2}}
\tablewidth{0pt}
\tablehead{
 & \multicolumn{9}{c}{$\delta<1$}\\
{\sc Object} & $m$\tablenotemark{a} & $\dot{m}$\tablenotemark{b} & 
$f$\tablenotemark{c} & $r_{\rm tr}$\tablenotemark{d} & 
$r_{\rm s}$\tablenotemark{d} & $kT_{\rm e}$\tablenotemark{e} & 
$\tau_{\rm c}$\tablenotemark{f} & $\tau_{\rm s}$\tablenotemark{f} & 
$\Gamma_{\rm fit}$\tablenotemark{g} & $\alpha_{\rm
  UV}$\tablenotemark{h} & $\alpha_{\rm ox}$\tablenotemark{i}
}
\startdata
BRI0103+0032 & $8.5{\times}10^9$ & 1.4 & 0.1 & 28 & 28 & 290 & 0.04 &
0.20 & 2.00 & -1.17 & 1.66\\
~&~&~&~&~&~&~&~&~&~&~&~\\
~& $7.5{\times}10^9$ & 1 & 0.1 & 19 & 9 & 200 & 0.07 & 0.37 & 1.96 &
-0.82 & 1.62\\
\tableline
PSS0248+1802 &\nodata&\nodata&\nodata&\nodata&\nodata&\nodata&\nodata&\nodata&\nodata&\nodata&\nodata\\
~&~&~&~&~&~&~&~&~&~&~&~\\
~& $1{\times}10^{10}$ & 1 & 0.1 & 15 & 7 & 105 & 0.18 & 0.71 & 2.01 &
-0.65 & 1.55\\
\tableline
BRI1033-0327 & $9{\times}10^9$ & 0.86 & 0.1 & 12 & 12 & 280 & 0.11 &
0.17 & 2.28 & -0.62 & 1.56\\
~ & $9{\times}10^9$ & 0.71 & 0.6 & 9 & 9 & 270 & 0.11 & 0.15 & 2.33 &
-0.48 & 1.55\\
~&~&~&~&~&~&~&~&~&~&~&~\\
~& \nodata&\nodata&\nodata&\nodata&\nodata&\nodata&\nodata&\nodata&\nodata&\nodata&\nodata\\
\tableline
PSS1057+4555 & \nodata&\nodata&\nodata&\nodata&\nodata&\nodata&\nodata&\nodata&\nodata&\nodata&\nodata\\
~&~&~&~&~&~&~&~&~&~&~&~\\
~& $1.5{\times}10^{10}$ & 1.3 & 0.1 & 15.5 & 4 & 310 & 0.04 & 0.25 &
1.86 & -0.82 & 1.73\\
~& $1.5{\times}10^{10}$ & 1 & 0.1 & 11.3 & 4 & 195 & 0.08 & 0.35 &
2.05 & -0.62 & 1.71\\
\tableline
PSS1317+3531 & $5{\times}10^9$ & 0.53 & 0.1 & 15 & 15 & 90 & 0.21 &
0.66 & 2.23 & -0.58 & 1.54\\
~& $3{\times}10^9$ & 0.98 & 0.1 & 15 & 15 & 110 & 0.17 & 0.54 & 2.23 &
-0.31 & 1.52\\ 
~& $5{\times}10^9$ & 0.48 & 0.15 & 12 & 12 & 90 & 0.25 & 0.61 & 2.36 &
-0.46 & 1.54\\
~&~&~&~&~&~&~&~&~&~&~&~\\
~& \nodata&\nodata&\nodata&\nodata&\nodata&\nodata&\nodata&\nodata&\nodata&\nodata&\nodata\\
\tableline
SDS1030+0524 & \nodata&\nodata&\nodata&\nodata&\nodata&\nodata&\nodata&\nodata&\nodata&\nodata&\nodata\\
~&~&~&~&~&~&~&~&~&~&~&~\\
~& $1{\times}10^{10}$ & 1 & 0.1 & 27 & 10 & 920 & 0.005 & 0.04 & 1.80
& -1.46 & 1.87\\
~& $6{\times}10^9$ & 0.93 & 0.1 & 17 & 4 & 500 & 0.02 & 0.13 & 1.80 &
-0.70 & 1.80\\
\tableline
SDS1204-0021 & \nodata&\nodata&\nodata&\nodata&\nodata&\nodata&\nodata&\nodata&\nodata&\nodata&\nodata\\
~&~&~&~&~&~&~&~&~&~&~&~\\
~& $8{\times}10^9$ & 0.88 & 0.1 & 16 & 4 & 425 & 0.02 & 0.16 & 1.83 &
-0.77 & 1.79\\
\tableline
PSS1443+2724 & $8{\times}10^9$ & 1.4 & 0.1 & 35 & 35 & 460 & 0.02 &
0.11 & 1.95 & -1.50 & 1.72\\
~&~&~&~&~&~&~&~&~&~&~&~\\
~& $7{\times}10^9$ & 0.65 & 0.1 & 15 & 5 & 200 & 0.07 & 0.39 & 1.95
& -0.74 & 1.65\\
~& $6{\times}10^9$ & 1 & 0.1 & 20 & 4 & 290 & 0.04 & 0.34 & 1.75 &
-0.82 & 1.65\\
\enddata

\tablenotetext{a}{Black hole mass in Solar masses, $M_{\odot}$}
\tablenotetext{b}{Accretion rate in Eddington units, $\dot{M}_{\rm Edd}$}
\tablenotetext{c}{Fraction of gravitational energy dissipated in the corona above 
the disk}
\tablenotetext{d}{Disk truncation radius,$r_{\rm tr}$, and the semi-minor axis of
the ellipsoidal flow, $r_{\rm s}$, in Schwarzschild units, $R_{\rm S}$}
\tablenotetext{e}{Plasma temperature in keV}
\tablenotetext{f}{Optical depth of corona, $\tau_c$, and semi-spherical flow,
$\tau_s$}
\tablenotetext{g}{X-ray photon index allowed to vary within the 1$\sigma$ confidence 
interval (we do not list $l_{\rm UV}$ from fits since its value was fixed at that observed)}
\tablenotetext{h}{UV index between 1450 and 2500 \AA\ in the rest frame}
\tablenotetext{i}{X-ray loudness}

\end{deluxetable}


\begin{thebibliography}{}

\bibitem[Band \& Malkan(1988)]{1988BAAS...20.1073B} Band, D.~L.~\& Malkan, 
M.~A.\ 1988, \baas, 20, 1073 

\bibitem[Bechtold et al.(1994)]{1994AJ....108..374B} Bechtold, J.~et al.\ 
1994a, \aj, 108, 374 

\bibitem[Bechtold et al.(1994)]{1994AJ....108..759B} Bechtold, J.~et al.\ 
1994b, \aj, 108, 759 


\bibitem[Bechtold et al.(2003)]{2003ApJ...588..119B} Bechtold, J.~et
al.\ 2003, \apj, 588, 119 (B03)

\bibitem[Becker et al.(2001)]{2001AJ....122.2850B} Becker, R.~H.~et al.\ 
2001, \aj, 122, 2850

\bibitem[Beloborodov(1999)]{1999ApJ...510L.123B} Beloborodov, A.~M.\ 1999a, 
\apjl, 510, L123 

\bibitem[Beloborodov(1999)]{1999hepa.conf..295B} Beloborodov, A.~M.\ 1999b, 
ASP Conf.~Ser.~161: High Energy Processes in Accreting Black Holes, 295 

\bibitem[Brandt et al.(2002)]{2002xsac.conf..235B} Brandt, W.~N., Vignali, 
C., Fan, X., Kaspi, S., \& Schneider, D.~P.\ 2002a, X-ray Spectroscopy of 
AGN with Chandra and XMM-Newton, 235

\bibitem[Brandt et al.(2002)]{2002ApJ...569L...5B} Brandt, W.~N.~et al.\ 
2002b, \apjl, 569, L5


\bibitem[Cappi et al.(1997)]{1997ApJ...478..492C} Cappi, M., Matsuoka, M., 
Comastri, A., Brinkmann, W., Elvis, M., Palumbo, G.~G.~C., \& Vignali, C.\ 
1997, \apj, 478, 492 

\bibitem[Chiang(2002)]{2002ApJ...572...79C} Chiang, J.\ 2002, \apj, 572, 79 

\bibitem[Chiang \& Blaes(2003)]{2003ApJ...586...97C} Chiang, J.~\& Blaes, 
O.\ 2003, \apj, 586, 97 

\bibitem[Collin-Souffrin, Czerny, Dumont, \& 
Zycki(1996)]{1996A&A...314..393C} Collin-Souffrin, S., Czerny, B., Dumont, 
A.-M., \& \.{Z}ycki, P.~T.\ 1996, \aap, 314, 393 

\bibitem[Coppi(1999)]{1999hepa.conf..375C} Coppi, P.~S.\ 1999, ASP 
Conf.~Ser.~161: High Energy Processes in Accreting Black Holes, 375 

\bibitem[Czerny \& Elvis(1987)]{1987ApJ...321..305C} Czerny, B.~\& Elvis, 
M.\ 1987, \apj, 321, 305 

\bibitem[Czerny et al.(2003)]{2003A&A...412..317C} Czerny, B., 
Niko{\l}ajuk, M., R{\' o}{\. z}a{\'n}ska, A., Dumont, A.-M., Loska, Z., \& 
Zycki, P.~T.\ 2003, \aap, 412, 317

\bibitem[Czerny(2003)]{2003agnc.conf...59C} Czerny, B.\ 2003, ASP 
Conf.~Ser.~290: Active Galactic Nuclei: From Central Engine to Host Galaxy, 
59 

\bibitem[Done, Madejski, Zycki(2000)]{2000ApJ...536..213D} Done, 
C., Madejski, G.~M., \.{Z}ycki, P.~T.\ 2000, \apj, 536, 213 

\bibitem[Done(2001)]{2001ncxa.conf..102D} Done, C.\ 2001, ASP 
Conf.~Ser.~251: New Century of X-ray Astronomy, 102


\bibitem[Esin, McClintock, \& Narayan(1997)]{1997ApJ...489..865E} Esin, 
A.~A., McClintock, J.~E., \& Narayan, R.\ 1997, \apj, 489, 865 

\bibitem[Fabian, Rees, Stella, \& White(1989)]{1989MNRAS.238..729F} Fabian, 
A.~C., Rees, M.~J., Stella, L., \& White, N.~E.\ 1989, \mnras, 238, 729 

\bibitem[Fan et al.(2001)]{2001AJ....121...31F} Fan, X.~et al.\ 2001, \aj, 
121, 31

\bibitem[Fiore et al.(1995)]{1995ApJ...449...74F} Fiore, F., Elvis, M., 
Siemiginowska, A., Wilkes, B.~J., McDowell, J.~C., \& Mathur, S.\ 1995, 
\apj, 449, 74 

\bibitem[Galeev, Rosner, \& Vaiana(1979)]{1979ApJ...229..318G} Galeev, 
A.~A., Rosner, R., \& Vaiana, G.~S.\ 1979, \apj, 229, 318 

\bibitem[George \& Fabian(1991)]{1991MNRAS.249..352G} George, I.~M.~\& 
Fabian, A.~C.\ 1991, \mnras, 249, 352 

\bibitem[Gierlinski et al.(1997)]{1997MNRAS.288..958G} Gierli{\'n}ski, M., 
Zdziarski, A.~A., Done, C., Johnson, W.~N., Ebisawa, K., Ueda, Y., Haardt, 
F., \& Phlips, B.~F.\ 1997, \mnras, 288, 958 

\bibitem[Gierli{\' n}ski et al.(1999)]{1999MNRAS.309..496G} Gierli{\' 
n}ski, M., Zdziarski, A.~A., Poutanen, J., Coppi, P.~S., Ebisawa, K., \& 
Johnson, W.~N.\ 1999, \mnras, 309, 496 

\bibitem[Gierli{\' n}ski \& Done(2004)]{2004MNRAS.349L...7G} Gierli{\' 
n}ski, M.~\& Done, C.\ 2004, \mnras, 349, L7

\bibitem[Giveon et al.(1999)]{1999MNRAS.306..637G} Giveon, U., Maoz, D., 
Kaspi, S., Netzer, H., \& Smith, P.~S.\ 1999, \mnras, 306, 637

\bibitem[Haardt \& Maraschi(1991)]{1991ApJ...380L..51H} Haardt, F.~\& 
Maraschi, L.\ 1991, \apjl, 380, L51 

\bibitem[Haardt \& Maraschi(1993)]{1993ApJ...413..507H} Haardt, F.~\& 
Maraschi, L.\ 1993, \apj, 413, 507 

\bibitem[Janiuk, Czerny, \& Madejski(2001)]{2001ApJ...557..408J} Janiuk, 
A., Czerny, B., \& Madejski, G.~M.\ 2001, \apj, 557, 408 

\bibitem[Janiuk, Czerny, Siemiginowska, \& 
Szczerba(2004)]{2004ApJ...602..595J} Janiuk, A., Czerny, B., Siemiginowska, 
A., \& Szczerba, R.\ 2004, \apj, 602, 595

\bibitem[Koratkar \& Blaes(1999)]{1999PASP..111....1K} Koratkar, A.~\& 
Blaes, O.\ 1999, \pasp, 111, 1 

\bibitem[Kuhn, Elvis, Bechtold, \& Elston(2001)]{2001ApJS..136..225K} Kuhn, 
O., Elvis, M., Bechtold, J., \& Elston, R.\ 2001, \apjs, 136, 225

\bibitem[Laor et al.(1997)]{1997ApJ...477...93L} Laor, A., Fiore, F., 
Elvis, M., Wilkes, B.~J., \& McDowell, J.~C.\ 1997, \apj, 477, 93 

\bibitem[Laor(1991)]{1991ApJ...376...90L} Laor, A.\ 1991, \apj, 376, 90 

\bibitem[Lawson \& Turner(1997)]{1997MNRAS.288..920L} Lawson, A.~J.~\& 
Turner, M.~J.~L.\ 1997, \mnras, 288, 920 

\bibitem[Lightman \& White(1988)]{1988ApJ...335...57L} Lightman, A.~P.~\& 
White, T.~R.\ 1988, \apj, 335, 57 

\bibitem[]{}
McClintock, J.~E.~\& Remillard, R.~A.\ 2003, astro-ph/0306213, to appear in
Compact Stellar X-ray Sources, eds. W.H.G. Lewin and M. van der Klis

\bibitem[Mineo et al.(2000)]{2000A&A...359..471M} Mineo, T.~et al.\ 2000, 
\aap, 359, 471 

\bibitem[Mushotzky, Done, \& Pounds(1993)]{1993ARA&A..31..717M} Mushotzky, 
R.~F., Done, C., \& Pounds, K.~A.\ 1993, \araa, 31, 717 

\bibitem[Nandra et al.(2000)]{2000ApJ...544..734N} Nandra, K., Le, T., 
George, I.~M., Edelson, R.~A., Mushotzky, R.~F., Peterson, B.~M., \& 
Turner, T.~J.\ 2000, \apj, 544, 734

\bibitem[Narayan, Mahadevan, \& Quataert(1998)]{1998tbha.conf..148N} 
Narayan, R., Mahadevan, R., \& Quataert, E.\ 1998, Theory of Black Hole 
Accretion Disks, 148 

\bibitem[Nayakshin(2000)]{2000ApJ...534..718N} Nayakshin, S.\ 2000a, \apj, 
534, 718 

\bibitem[Nayakshin, Kazanas, \& Kallman(2000)]{2000ApJ...537..833N} 
Nayakshin, S., Kazanas, D., \& Kallman, T.~R.\ 2000b, \apj, 537, 833 

\bibitem[Page et al.(2003)]{2003MNRAS.338.1004P} Page, K.~L., Turner, 
M.~J.~L., Reeves, J.~N., O'Brien, P.~T., \& Sembay, S.\ 2003, \mnras, 338, 
1004 

\bibitem[Pentericci et al.(2003)]{2003A&A...410...75P} Pentericci, L.~et 
al.\ 2003, \aap, 410, 75

\bibitem[Poutanen, Krolik, \& Ryde(1997)]{1997MNRAS.292L..21P} Poutanen, 
J., Krolik, J.~H., \& Ryde, F.\ 1997, \mnras, 292, L21 

\bibitem[Reeves \& Turner(2000)]{2000MNRAS.316..234R} Reeves, J.~N.~\& 
Turner, M.~J.~L.\ 2000, \mnras, 316, 234 

\bibitem[R{\' o}{\. z}a{\' N}ska \& Czerny(2000)]{2000A&A...360.1170R} R{\' 
o}{\. z}a{\' N}ska, A.~\& Czerny, B.\ 2000, \aap, 360, 1170 

\bibitem[Shakura \& Sunyaev(1973)]{1973A&A....24..337S} Shakura, N.~I.~\& 
Sunyaev, R.~A.\ 1973, \aap, 24, 337 

\bibitem[Shapiro, Lightman, \& Eardley(1976)]{1976ApJ...204..187S} Shapiro, 
S.~L., Lightman, A.~P., \& Eardley, D.~M.\ 1976, \apj, 204, 187 

\bibitem[Shields(1978)]{1978Natur.272..706S} Shields, G.~A.\ 1978, \nat, 
272, 706 

\bibitem[Siebert et al.(1996)]{1996A&A...307....8S} Siebert, J., Matsuoka, 
M., Brinkmann, W., Cappi, M., Mihara, T., \& Takahashi, T.\ 1996, \aap, 
307, 8 

\bibitem[Siemiginowska et al.(1995)]{1995ApJ...454...77S} Siemiginowska, 
A., Kuhn, O., Elvis, M., Fiore, F., McDowell, J., \& Wilkes, B.~J.\ 1995, 
\apj, 454, 77 

\bibitem[]{}
Sobolewska, M.~A., Siemiginowska, A., \& \.{Z}ycki, P.~T.\ 2004,
accepted for publication in \apj ~(Paper II)

\bibitem[Stern et al.(1995)]{1995ApJ...449L..13S} Stern, B.~E., Poutanen, 
J., Svensson, R., Sikora, M., \& Begelman, M.~C.\ 1995, \apjl, 449, L13 

\bibitem[Stern et al.(2000)]{2000AJ....119.1526S} Stern, D., Djorgovski, 
S.~G., Perley, R.~A., de Carvalho, R.~R., \& Wall, J.~V.\ 2000, \aj, 119, 
1526 

\bibitem[Sun \& Malkan(1989)]{1989ApJ...346...68S} Sun, W.~\& Malkan, 
M.~A.\ 1989, \apj, 346, 68 

\bibitem[Tripp, Bechtold, \& Green(1994)]{1994ApJ...433..533T} Tripp, 
T.~M., Bechtold, J., \& Green, R.~F.\ 1994, \apj, 433, 533 

\bibitem[Vignali et al.(2003)]{2003AJ....125..418V} Vignali, C., Brandt, 
W.~N., Schneider, D.~P., Garmire, G.~P., \& Kaspi, S.\ 2003a, \aj, 125, 418 (V03)


\bibitem[Vignali et al.(2003)]{2003AJ....125.2876V} Vignali, C.~et al.\ 
2003b, \aj, 125, 2876 

\bibitem[Wilson \& Done(2001)]{2001MNRAS.325..167W} Wilson, C.~D.~\& Done, 
C.\ 2001, \mnras, 325, 167 

\bibitem[Worrall, Tananbaum, Giommi, \& 
Zamorani(1987)]{1987ApJ...313..596W} Worrall, D.~M., Tananbaum, H., Giommi, 
P., \& Zamorani, G.\ 1987, \apj, 313, 596 


\bibitem[Zamorani et al.(1981)]{1981ApJ...245..357Z} Zamorani, G.~et al.\ 
1981, \apj, 245, 357 

\bibitem[Zdziarski, Lubinski, \& Smith(1999)]{1999MNRAS.303L..11Z} 
Zdziarski, A.~A., Lubi{\'n}ski, P., \& Smith, D.~A.\ 1999, \mnras, 303, L11 

\bibitem[Zdziarski et al.(1998)]{1998MNRAS.301..435Z} Zdziarski, A.~A., 
Poutanen, J., Miko{\l}ajewska, J., Gierli{\'n}ski, M., Ebisawa, K., \& Johnson, 
W.~N.\ 1998, \mnras, 301, 435 

\bibitem[Zdziarski, Poutanen, \& Johnson(2000)]{2000ApJ...542..703Z} 
Zdziarski, A.~A., Poutanen, J., \& Johnson, W.~N.\ 2000, \apj, 542, 703 

\bibitem[Zycki \& Czerny(1994)]{1994MNRAS.266..653Z} \.{Z}ycki, P.~T.~\& 
Czerny, B.\ 1994, \mnras, 266, 653 

\bibitem[{\. Z}ycki, Done, \& Smith(2001)]{2001MNRAS.326.1367Z} {\. Z}ycki, 
P.~T., Done, C., \& Smith, D.~A.\ 2001, \mnras, 326, 1367 

\end{thebibliography}
\end{document}